\documentclass[journal]{IEEEtran}

\usepackage{graphicx}
\usepackage{caption}

\usepackage{algorithm}
\usepackage[noend]{algpseudocode}

\usepackage{multicol}
\usepackage{amsmath,amssymb}
\usepackage{amsfonts}
\usepackage{textcomp}
\usepackage{psfrag}
\usepackage{multimedia}

\newcommand{\E}[2]{\mathbb{E}_{#1}\left[#2\right]}

\newcommand{\vect}[1]{\ensuremath{\boldsymbol{\mathrm{#1}}}}
\newtheorem{theorem}{Theorem}
\newtheorem{Proposition}{Proposition}

\newtheorem{Corollary}{Corollary}
\newtheorem{Lemma}{Lemma}
\newtheorem{Definition}{Definition}
\newtheorem{Remark}{Remark}

\newtheorem{Limitation}{Fundamental Limitation}

\usepackage[usenames,dvipsnames]{color}
\definecolor{wheat}{rgb}{0.96,0.87,0.70}
\definecolor{mario}{rgb}{0.8,0.8,1}
\definecolor{seb}{rgb}{0.8,1,0.8}
\definecolor{myGreen}{rgb}{0.,0.8,0.0}

\newcommand {\matr}[2]{\left[\begin{array}{#1}#2\end{array}\right]}

\newcounter{lastnote}

\usepackage{tikz}
\usetikzlibrary{calc,arrows,positioning}

\newcommand{\change}[1]{{\color{blue}#1}}

\begin{document} 

\title{Safe Reinforcement Learning Using Robust MPC} 

\author
{Mario Zanon and S\'ebastien Gros
	\thanks{M. Zanon is with the IMT School for Advanced Studies Lucca, Italy. }
	\thanks{S. Gros is with the Department of Engineering Cybernetics, NTNU, Norway.}
}

\IEEEtitleabstractindextext{
	\begin{abstract}
		Reinforcement Learning (RL) has recently impressed the world with stunning results in various applications. While the potential of RL is now well-established, many critical aspects still need to be tackled, including safety and stability issues. These issues, while secondary for the RL community, are central to the control community which has been widely investigating them. Model Predictive Control (MPC) is one of the most successful control techniques because, among others, of its ability to provide such guarantees even for uncertain constrained systems. 
		Since MPC is an optimization-based technique, optimality has also often been claimed. Unfortunately, the performance of MPC is highly dependent on the accuracy of the model used for predictions. In this paper, we propose to combine RL and MPC in order to exploit the advantages of both and, therefore, obtain a controller which is optimal and safe. We illustrate the results with two numerical examples in simulations.

	\end{abstract}
	
	\begin{IEEEkeywords}
		Reinforcement Learning, Robust Model Predictive Control, safe policies
	\end{IEEEkeywords}
}

\maketitle

\IEEEdisplaynontitleabstractindextext

\IEEEpeerreviewmaketitle

\section{Introduction}
Reinforcement Learning (RL) is a technique for solving problems involving Markov Decision Processes (MDP)~\cite{Sutton1998}. In RL, rather than modeled state transition probabilities, samples and observed costs (or rewards) are used.
RL algorithms enabled computers beating Chess and Go masters \cite{Silver2016}, and robots learning to walk or fly without supervision \cite{Wang2012,Abbeel2007}.

For each state $\vect{s}$ the optimal action $\vect{a}$ is computed as the the optimal feedback policy $\vect\pi(\vect s)$ for the real system either directly (policy search methods)~\cite{Sutton1999,Silver2014} or indirectly (SARSA, $Q$-learning)~\cite{Watkins1989}. 
In the latter, the optimal policy $\vect\pi(\vect s)$ is indirectly obtained as the minimizer of the so-called action-value function $Q(\vect s,\vect a)$ over the action or input $\vect a$. In both cases, either $\vect \pi$ or $Q$ are typically approximated by a function approximator: Deep Neural Network (DNN) are very commonly used for that purpose in recent applications.

While RL has demonstrated in practice a huge potential, properties that are typically expected from a controller, such as, e.g., some form of stability and safety, are hard to guarantee, especially when relying on a DNN as a function approximator. In general, any safety-enforcing approach requires either (a) to collect data from the real system thus incurring catastrophic events, (b) to use a model of the system to generate a sufficient amount of stochastic simulations, or (c) to model the uncertainty underlying the system, as we will detail later. Some approaches have been developed in order to guarantee some form of safety: see, e.g., the excellent survey in~\cite{Garcia2015} and references therein. Most approaches, however, do not strictly guarantee that a given set of constraints is never violated, but rather that violations are rare events. Some approaches propose to project the action resulting from a DNN onto a safe set: in~\cite{Dalal2018} each constraint is approximated as a ReLu or an additional cost and in~\cite{Pham2018} a QP is used. In both approaches the nominal prediction is used and uncertainty is neglected in the predictions, similarly to the approach of~\cite{Gros2018}. Projection approaches have been analyzed in the general case in~\cite{Gros2019d}.

The combination of learning and control techniques has been proposed in, e.g.,~\cite{Koller2018,Aswani2013,Ostafew2016,Berkenkamp2017,Murray2018}. The combination of RL and the linear quadratic regulator has been presented in~\cite{Lewis2009,Lewis2012}. To the best of our knowledge, \cite{Gros2018,Zanon2019,Amos2018} are the first works proposing to use NMPC as a function approximator in RL. While strategies for providing some form of safety have been developed~\cite{Garcia2015}, to the best of the authors' knowledge, none of these approaches is able to strictly satisfy some set of constraints at all time. Rather, constraint violation is strongly penalized in~\cite{Gros2018} and some of the approaches in~\cite{Garcia2015}.
The only contribution providing robust constraint satisfaction guarantees is~\cite{Dean2019}, where a linear feedback policy is learned.

In this paper we propose an RL formulation based on MPC which addresses the issue of safety, which we did not rigorously enforce in~\cite{Gros2018,Zanon2019}. 
We summarize next two existing approaches and the scheme we propose, which combines them: 
\paragraph{Robust MPC, Business-as-usual} a low-dimensional computationally tractable uncertainty set is first identified and then used to formulate a robust MPC problem, see~\eqref{eq:robust_mpc}.
\paragraph{RL, Business-as-usual} penalizing violations with a suitably high cost lets the optimization procedure yield a policy which tends to not violate the constraints. Safety is typically not strictly guaranteed and only few results provide weak guarantees. 
\paragraph{Safe RL MPC} In the approach we propose, a robust MPC problem is formulated similarly to (a). Similarly to (b), RL updates the parametrization of the robust MPC scheme, and of the safety constraint to reduce conservatism while preserving safety. 

Safe RL-MPC is based on the approach first advocated in~\cite{Gros2018,Zanon2019}. The scheme can be seen from two alternative points of view: (a) MPC is used as a function approximator within RL in order to provide safety and stability guarantees; and (b) RL is used in order to tune the MPC parameters, thus improving closed-loop performance in a data-driven fashion. 
Since safety is fundamental not only during exploitation, but also during exploration, we also address the issue of guaranteeing constraint satisfaction during this phase. 

Another important contribution of this paper is the development of an efficient way to deal with the enormous amount of data typically collected by autonomous systems. In particular, the introduction of a nominal (potentially inaccurate) linear model allows one to significantly reduce the amount of stored data. Further efficiency is obtained by exploiting convexity and using a low-dimensional approximation of the uncertainty set.

We first present RL at a conceptual level and propose adaptations in order to make RL applicable to the safety-enforcing setup. The main issue to be tackled is related to the safety constraints, which need to be enforced when updating the function approximator parameter. We formulate the update by resorting to a constrained optimization problem, similarly to what has been proposed in~\cite{Zanon2019}. The proposed safe RL can be directly applied to $Q$-learning, but actor-critic techniques require some adaptation when the input space is continuous and restricted by safety requirements as discussed in~\cite{Gros2019e}.

\paragraph*{Contributions} This paper proposes an approach to combine MPC and RL so as to guarantee safety. In order to limit the complexity of the algorithm, we (a) rely on a linear system with bounded disturbances for predictions, and (b) propose an ad-hoc formulation of the robust constraint satisfaction which relies on the linear model to greatly reduce the amount of data to be stored. Finally, adaptations to the standard RL algorithms to guarantee that the parameter update does not jeopardize safety are introduced.

The paper is structured as follows. We introduce the problem of safe RL in general terms in Section~\ref{sec:safe_rl}, while the rest of the paper specializes on the case of linear systems. We propose a tailored function approximator based on robust MPC in Section~\ref{sec:rmpc} and discuss the efficient use of data in Section~\ref{sec:data}. The necessary modifications to the standard RL algorithms are proposed in Section~\ref{sec:rl} and the whole framework is tested in simulations in Section~\ref{sec:simulations}. Conclusions and an outline for future research are given in Section~\ref{sec:conclusions}.

\paragraph*{Notation}
$a$ is scalar, $\vect{a}\in\mathbb{R}^{n_{\vect{a}}}$ is a vector with components $\vect{a}_i$, $A$ is a matrix with rows $A_i$ and $\vect{A}$, $\mathcal{A}$ are sets. For any set, $|\cdot|$ defines its cardinality. 
For any function $\vect{f}(\vect{x})$, we define $\vect{f}(\vect{X}):=\{ \, \vect{f}(\vect{x}) \, | \, \vect{x}\in\vect{X}  \,\}$ and denote $\vect{f}(\vect{x})\leq 0, \ \forall \, \vect{x}\in \vect{X}$ as $\vect{f}(\vect{X})\leq 0$. The only exceptions are $J$, $V$, $Q$ which are scalar functions, but denoted by capital letters in the literature.

\begin{figure*}
	\begin{center}
		\begin{tikzpicture}[scale=0.9, every node/.style={scale=0.9}]
		\tikzstyle{block} = [draw, rectangle, minimum height=1cm, minimum width=0.1cm]
		\tikzstyle{blockBig} = [draw, rectangle, dotted, minimum height=3.2cm, minimum width=0.1cm]
		\tikzstyle{block2} = [draw, rectangle, fill=white, minimum height=0.1cm, minimum width=0.1cm]
		\node [block,text width=3cm, align=center] (noise) {$\vect{s}_+ - A\vect{s}- B\vect{a} - \vect{b}$};
		\node [above right = -0.3cm and 0.5cm of noise,text width=1cm, align=center,anchor=center] {$\vect{w}$};
		\node [block,right = 1cm of noise,text width=2.3cm, align=center] (sdc) {$\mathrm{Conv}(\{\vect{w},\mathcal{\bar W}\})$};
		
		\coordinate [right = 0.5cm of noise] (noise1) {};
		\coordinate [below = 0.6cm of noise1] (noise2) {};
		\coordinate [right = 3.3cm of noise2] (noise3) {};
		\coordinate [above = 0.35cm of noise3] (noise4) {};
		\coordinate [right = 0.5cm of noise4] (noise5) {};
		\draw [-latex] (noise.east) -- (sdc);
		\draw [-] (noise.east) -- (noise1);
		\draw [-] (noise1) -- (noise2);
		\draw [-] (noise2) -- (noise3);
		\draw [-] (noise3) -- (noise4);
		\draw [-latex] (noise4) -- (noise5);
		\node [above right = -0.3cm and 0.3cm of sdc,text width=1cm, align=center,anchor=center] {$\mathcal{\bar W}$};
		\node [block,right = 1cm of sdc,text width=2.5cm, align=center] (rl) {Safe RL \\ \eqref{eq:safe_rl_problem_sampled} $\equiv$ \eqref{eq:safe_rl_sampled} };
		\node [above right = -0.3cm and 0.3cm of rl,text width=1cm, align=center,anchor=center] {$\vect{\theta}$};
		\draw [-latex] (sdc.east) -- (rl.west);
		\node [block,right = 1cm of rl,text width=2cm, align=center] (mpc) {Robust MPC \\ \eqref{eq:robust_mpc} or \eqref{eq:exploration}};
		\node [above right = -0.3cm and 0.5cm of mpc,anchor=center] {$\vect{a}$};
		\node [below right = -0.2cm and 0.5cm of mpc,anchor=center] {$\vect{\pi}_{\vect{\theta}}(\vect{s})$};
		\draw [-latex] (rl.east) -- (mpc.west);
		\node [block,right = 1cm of mpc,text width=2cm, align=center] (system) {System \\ \eqref{eq:State:Transition}};
		\node [above right = -0.3cm and 0.5cm of system,anchor=center] {$\vect{s}_+$};
		\draw [-latex] (mpc.east) -- (system.west);
		\coordinate [right = 1cm of system] (system1);
		\coordinate [above = 1cm of system1] (system2);
		\draw [-] (system.east) -- (system1);
		\draw [-] (system1) -- (system2);
		\node [block2,left = 1.7cm of system2,text width=0.5cm, align=center] (z) {$z^{-1}$};
		\draw [-latex] (system2) -- (z);
		\coordinate (z1) at (mpc |- z);
		\node [below right = 0.2cm and 0.2cm of z1,anchor=center] {$\vect{s}$};
		\draw [-] (z.west) -- (z1);
		\draw [-latex] (z1) -- (mpc.north);
		\coordinate [right = 0.2cm of noise.north] (z3);
		\coordinate (z2) at (z3 |- z);
		\node [below right = 0.2cm and 0.2cm of z2,anchor=center] (sinnoise) {$\vect{s}$};
		\draw [-] (z1) -- (z2);
		\draw [-latex] (z2) -- (z3);
		\coordinate [right = 0.3cm of mpc] (a1);
		\coordinate [above = 1.2cm of a1] (a2);
		\coordinate [left = 0.2cm of noise.north] (a4);
		\coordinate (a3) at (a4 |- a2);
		\draw [-] (a1) -- (a2);
		\draw [-] (a2) -- (a3);
		\draw [-latex] (a3) -- (a4);
		\node [left = 0.65cm of sinnoise,anchor=center] {$\vect{a}$};
		
		\coordinate [below = 0.6cm of system] (ell1);
		\node [above left = 0.2cm and 0.5cm of ell1,anchor=center] {$\ell(\vect{s},\vect{a})$};
		\coordinate [left = 0.2cm of rl.south] (ell3);
		\coordinate (ell2) at (ell3 |- ell1);
		\draw [-] (system.south) -- (ell1);
		\draw [-] (ell1) -- (ell2);
		\draw [-latex] (ell2) -- (ell3);
		
		\coordinate [below = 0.5cm of mpc] (q1);
		\node [above left = 0.2cm and 1.3cm of q1,anchor=center] {$Q_{\vect{\theta}}, \vect{\pi}_{\vect{\theta}}, \nabla Q_{\vect{\theta}},\nabla \vect{\pi}_{\vect{\theta}}$};
		\coordinate [right = 0.2cm of rl.south] (q3);
		\coordinate (q2) at (q3 |- q1);
		\draw [-] (mpc.south) -- (q1);
		\draw [-] (q1) -- (q2);
		\draw [-latex] (q2) -- (q3);
			
			\node [blockBig,text width=7.5cm, align=center] at (2.1,-0.1) (secBlock1) {};
			\node [below = -0.4cm of secBlock1] {Section~\ref{sec:data}};
			\node [blockBig,right = 0.05cm of secBlock1, text width=3.4cm, align=center] (secBlock2) {};
			\node [below = -0.4cm of secBlock2] {Section~\ref{sec:rl}};
			\node [blockBig,right = 0.05cm of secBlock2, text width=3.5cm, align=center] (secBlock3) {};
			\node [below = -0.4cm of secBlock3] {Section~\ref{sec:rmpc}};
		\end{tikzpicture}
	\end{center}
	\caption{Schematics of the proposed setup: data are used to construct the SDC based on $\mathcal{\bar W}$ and to evaluate the cost in~\eqref{eq:rl_problem_sampled}. This cost depends on $Q_{\vect{\theta}},V_{\vect{\theta}}$ obtained from MPC, and $\ell$. MPC controls the system. The signal toggling between exploitation and exploration is omitted to avoid confusion, and is a signal sent from RL to switch between MPC~\eqref{eq:robust_mpc} and~\eqref{eq:exploration}.}
	\label{fig:rl-mpc}
\end{figure*}
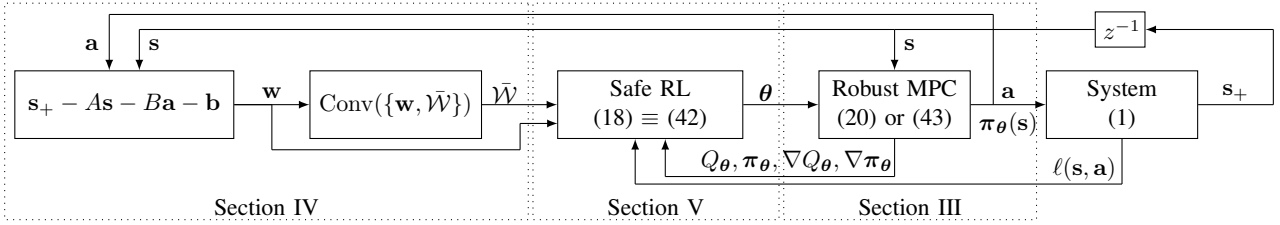

\section{Background and Safe RL Formulation}
\label{sec:safe_rl}

In this paper, we consider real system dynamics described as a Markov Process (MP) with continuous state $\vect{s}$ and action $\vect{a}$, with state transitions $\vect{s},\vect{a}\to \vect{s}_{+}$ having the underlying conditional probability density
\begin{align}
\label{eq:State:Transition}
\mathbb{P}\left[\vect{s}_{+}\,|\,\vect{s},\vect{a}\right].
\end{align}
We furthermore consider a deterministic policy delivering the control input as
$
	\vect a = \vect{\pi}\left(\vect s\right),
$ 
resulting in state distribution $\tau^{\vect{\pi}}$. 
The RL problem then reads as
\begin{align}
\label{eq:rl_problem}
\vect{\pi}_\star:=\arg\min_{\vect{\pi}}\ J(\vect{\pi}) & := \E{\tau^{\vect{\pi}}}{\sum_{k=0}^\infty \gamma^k \ell\left (\vect{s}_k,\vect{\pi}\left (\vect{s}_k\right )\right )},
\end{align}
where $\ell$ is called stage cost in optimal control and $-\ell$ instantaneous reward in RL. The scalar $\gamma$ is a discount factor, typically smaller than $1$ in RL, and $1$ in MPC. Note that in~\eqref{eq:rl_problem} we provide the definition for non-episodic settings, but the developments of the paper readily apply to episodic settings. 
The value function $V_\star(\vect{s})$ is the optimal cost, obtained by applying the optimal policy $\vect{\pi}_\star$, i.e.,
\begin{align}
V_\star(\vect s_0) &=  \E{\tau^{\vect{\pi}_\star}}{\left.\sum_{k=0}^\infty \gamma^k \ell\left (\vect{s}_k,\vect{\pi}_\star\left (\vect{s}_k\right )\right )\,\right|\, \vect s_0 };
\end{align} 
and the Bellman Equation defines the action-value function
\begin{align}
\label{eq:bellman_error}
Q_\star(\vect{s},\vect{a}) := \ell(\vect{s},\vect{a}) + \gamma \E{}{ \, V_\star(\vect{s_+})\,|\,\vect{s},\vect{a} \, },
\end{align}
where the expectation is taken over the state transition~\eqref{eq:State:Transition}. 
The various forms of RL use parametric approximations $V_{\theta}$, $Q_{\theta}$, $\vect{\pi}_{\theta}$ of $V_\star$, $Q_\star$, $\vect{\pi}_\star$ in order to find the parameter $\vect{\theta}^\star$ which (approximately) solves~\eqref{eq:rl_problem} either directly or by approximately solving~\eqref{eq:bellman_error} in a sampled-based fashion. For both approaches we summarize this as
\begin{align}
	\label{eq:rl_problem_sampled}
	\vect{\theta}^\star := \min_{\vect{\theta}}&\ \sum_{k=0}^n \psi(\vect{s}_{k+1},\vect{s}_k,\vect{a}_k,\vect{\theta}),
\end{align}
where function $\psi$ depends on the specific algorithm, e.g.,
\begin{align}
\label{eq:q_learning}
\psi(\vect{s}_{k+1},\vect{s}_k,\vect{a}_k,\vect{\theta}) = \left ( \ell(\vect{s}_k,\vect{a}_k) + \gamma V_{\vect{\theta}_k}(\vect{s}_{k+1}) - Q_{\vect{\theta}}(\vect{s}_k,\vect{a}_k) \right )^2
\end{align}
and $n=1$ in recursive $Q$ learning formulations; and
\begin{align}
\label{eq:policy_gradient}
\E{\tau^{\vect{\pi}_{\theta}}}{\psi(\vect{s}_{k+1},\vect{s}_k,\vect{a}_k,\vect{\theta})} = J(\vect{\pi}_{\vect{\theta}})
\end{align}
in policy gradient approaches~\cite{Sutton1998}. 
To optimize performance, the system ought to be controlled by using the best available policy $\vect{\pi}_{\vect{\theta}}$, this is referred to as \emph{exploitation}. However, in order for Problem~\eqref{eq:rl_problem_sampled} to be well-posed in general, it is necessary to also collect data by deviating from $\vect{\pi}_{\vect{\theta}}$ and implementing a different policy $\vect{\pi}^\mathrm{e}$, this is referred to as \emph{exploration}.

Among others, one of the main difficulties related with RL is safety enforcement, e.g., collision avoidance. 
In this paper we define safety through a set of constraints 
\begin{align}
	\vect{\xi}(\vect{s},\vect{\hat \pi}(\vect{s}))\leq 0, && \forall \ \vect{s} \in \mathrm{supp}\left (\tau^{\vect{\hat \pi}}\right ), \label{eq:SafetyConst}
\end{align}
where we note $\mathrm{supp}\left (\tau^{\vect{\hat \pi}}\right )$ the support of the distribution of the MP~\eqref{eq:State:Transition} subject to policy $\hat{\vect \pi}$. Ideally, condition \eqref{eq:SafetyConst} should be satisfied at all times with unitary probability, both during exploitation, i.e., $\vect{\hat \pi}=\vect{\pi}_{\vect{\theta}}$, and exploration, i.e., $\vect{\hat \pi}=\vect{\pi}^\mathrm{e}\neq\vect{\pi}_{\vect{\theta}}$. Note that~\eqref{eq:SafetyConst} can only hold if process~\eqref{eq:State:Transition} has bounded support.

Enforcing~\eqref{eq:SafetyConst} poses two major challenges: (i) either the support of~\eqref{eq:State:Transition} or the support of $\tau^{\vect{\hat \pi}}$ must be known or estimated; (ii) given knowledge on either supports, a policy satisfying~\eqref{eq:SafetyConst} must be designed.

Problem (i) is fundamental, since one can never have the guarantee of being able to observe the full support of~\eqref{eq:State:Transition}. Arguably, a reasonable approach can be to approximate the support based on the information extracted from the available samples, or on a prior, or on both. We assume that collected data is informative such that, in the limit for an infinite amount of data, the support is reconstructed exactly for policy $\vect{\pi}_{\vect{\theta}}$.
A theoretical justification of this is beyond the scope of this paper. 
In order to construct an approximation of the MP support, we first define the dispersion set confining the state transitions as:
\begin{align}
	\vect{S}_+\left(\vect{s},\vect{a}\right) = \{\, \vect{s}_+ \,\,|\,\, \mathbb{P}\left[\vect{s}_{+}\,|\,\vect{s},\vect{a}\right] >0 \,\}.
\end{align}
Since $\vect{S}_+$ is not known a priori, we introduce a parametrized approximation $\vect{\hat S}_+$ based on function $\vect{g}_{\vect{\theta}}$ given by: 
\begin{align}
	\label{eq:param_disp_set}
	\vect{\hat S}_+(\vect{s},\vect{a},\vect{\theta}) := \{ \, \vect{s}_+ \, | \, \vect{g}_{\vect{\theta}}(\vect{s}_+,\vect{s},\vect{a}) \leq 0  \, \}.
\end{align}
In order to enforce safety, $\vect{\hat S}_+$ must be an outer approximation of set $\vect{S}_+$, i.e., $\vect{\theta}$ must be chosen such that:
\begin{align}
	\label{eq:sdc}
	\vect{\hat S}_+(\vect{s},\vect{a},\vect{\theta}) \supseteq \vect{S}_+(\vect{s},\vect{a}), && \forall \ \vect{s},\vect{a}.
\end{align}
We label this condition \emph{Safe-Design Constraint} (SDC), since it restricts the values that $\vect{\theta}$ can take based on safety concerns.
	In order to discuss safety in mathematically simple terms, let us introduce the worst-case mass of \eqref{eq:State:Transition} outside of $\vect{\hat S}_+$, defined as:
\begin{align}
\chi\left(\vect\theta\right) = \sup_{\vect s,\vect a}\mathbb E\left[\left.\vect s_+ \notin \vect{\hat S}_+(\vect{s},\vect{a},\vect{\theta})\,\right |\, \vect s,\vect a \right] \in [0,1],
\end{align}	
where the expected value is taken over \eqref{eq:State:Transition}. For a given $\vect \theta$, \eqref{eq:sdc} is ensured if $\chi\left(\vect\theta\right)=0$. However, $\chi\left(\vect\theta\right) $ is known only if we assume the real system dynamics \eqref{eq:State:Transition} are known. Otherwise, in the Bayesian context, $\chi\left(\vect\theta\right)$ ought to be treated as a random variable, reflecting our imperfect knowledge of it, conditioned on the current data $\mathcal D$ (i.e., knowledge) we have of the system. 
We therefore consider
\begin{align}
\eta\left(\mathcal D\right) = \mathbb P\left[\,\chi\left(\vect\theta\right)=0\,|\, \mathcal D\,\right],
\end{align}
%
which provides a formal definition of the probability that $\vect{\hat S}_+$ is capturing the real system dispersion provided the data $\mathcal D$. While computing $\eta$ is difficult in the general case, its definition allows us to discuss in rigorous terms the fundamental limitation provided next.

	\begin{Limitation}
		\label{lim:safety}
		For any $\mathcal D$, if $\vect{\hat S}_+\subset \mathbb{R}^{n_{\vect{s}}}$, 
		it is impossible to guarantee that 
		\begin{align}
		\label{eq:WeAreGod}
		\eta\left(\mathcal D\right) = 1
		\end{align}
		without introducing any additional assumption on \eqref{eq:State:Transition}.
	\end{Limitation}
	
	In other words, given a finite set of samples and, possibly, prior but not absolute knowledge of the system, one does not have enough information to construct a set containing all future samples, unless this set is $\mathbb{R}^{n_{\vect{s}}}$. This is a fundamental limitation of robust constraint satisfaction and, therefore, it is independent of the proposed approach. In the following, we will therefore discuss $\eta$-safety, underlining that \eqref{eq:WeAreGod} cannot be achieved in practice.
	
	Arguably, by introducing additional assumptions restricting the function space to which~\eqref{eq:State:Transition} can belong, one can envision overcoming this limitation. The investigation of this topic is beyond the scope of this paper and will be the subject of future research.
		
In order to formulate some form of SDC with limited information, we introduce the sample-based form of~\eqref{eq:sdc} as
\begin{align}
	\label{eq:sdc_sampled}
	\vect{s}_{k+1} \in \vect{\hat S}_+(\vect{s}_k,\vect{a}_k,\vect{\theta}), && \forall \ k.
\end{align}

For problem (ii), the main challenge is to find values of $\vect{\theta}$ such that the policy $\vect{\pi}_{\vect{\theta}}$ strictly satisfies the safety constraints. 

	We define safety based on the set dispersion propagation under policy $\vect{\pi}_{\vect{\theta}}$, which reads as
	\begin{align}
	\label{eq:SetDispersion}
	\vect{S}_{k+1}^{\vect{\pi}_{\vect{\theta}}} &:= \vect{\hat S}_+\left(\vect{S}_{k}^{\vect{\pi}_{\vect{\theta}}},\vect{\pi}_{\vect{\theta}}\left (\vect{S}_{k}^{\vect{\pi}_{\vect{\theta}}}\right ), \vect{\theta} \right), \qquad \vect{S}_0^{\vect{\pi}_{\vect{\theta}}} = \vect{s}_0.
	\end{align}
	\begin{Definition}[$\eta$-safe Policy]
		\label{def:safety}
		For a given data set $\mathcal D$ and a given set of initial conditions $\mathcal{S}_0$, a policy $\vect{\pi}_{\vect{\theta}}$ is labeled as $\eta$-safe for initial states $\vect{s}_0\in\mathcal{S}_0$ 
		if it satisfies 
		\begin{align}
		\vect{\xi}(\vect{S}_k^{\vect{\pi}_{\vect{\theta}}},\vect{\pi}_{\vect{\theta}}(\vect{S}_k^{\vect{\pi}_{\vect{\theta}}}))\leq 0, && \forall \, k \geq 0.
		\end{align}
	\end{Definition}
	In general, there can exist initial states for which a safe policy cannot exist~\cite{Rawlings2012b}, and Definition~\ref{def:safety} characterizes policies that preserve safety for a given initial state.

Providing safety guarantees is arguably an open problem when using DNNs as function approximators. 
However, this problem has been studied in control theory and one successful design technique is robust MPC~\cite{Chisci2001,Mayne2014,Mayne2015}. 
Therefore, instead of building the function approximations based on the commonly used DNN approaches, we will use robust MPC, within an extended version of the RL-MPC scheme proposed in~\cite{Gros2018,Zanon2019}. 
Note that it has been proven in~\cite{Gros2018} that the optimal policy $\vect{\pi}$, value and action-value functions $V_\star$ and $Q_\star$ can be recovered exactly by function approximations based on MPC, provided that their parametrization is rich enough, even in case the model used in MPC is different from~\eqref{eq:State:Transition}.

In order to guarantee safety, robust MPC would ideally rely on the propagation of the dispersion set~\eqref{eq:SetDispersion}. Unfortunately, this poses severe computational challenges and an auxiliary (time-varying) policy $\vect{\pi}^{\mathrm{MPC}}_{\vect{\theta},k}$ is preferred in order to recover a computationally tractable formulation~\cite{Mayne2014}. Policy $\vect{\pi}^{\mathrm{MPC}}_{\vect{\theta},k}$ is typically selected as an open-loop input profile corrected by an affine feedback, see Section~\ref{sec:constr_tightening}.
We remark that MPC delivers $\vect{\pi}_{\vect{\theta}}=\vect{\pi}^{\mathrm{MPC}}_{\vect{\theta},0}$. 
By construction, robust MPC delivers a policy $\vect{\pi}_{\vect{\theta}}$ which satisfies constraints $\vect{\xi}$ at all future times, provided that the SDC~\eqref{eq:sdc_sampled} holds for all $k$. Even though the dispersion set propagation $\vect{S}_{k}^{\vect{\pi}^{\mathrm{MPC}}_{\vect{\theta},k}}$ is computed based on $\vect{\pi}^{\mathrm{MPC}}_{\vect{\theta},k}$, the constraints are guaranteed to hold also for $\vect{S}_{k}^{\vect{\pi}_{\vect{\theta}}}$.

Since the shape of $\vect{\hat S}_+$ impacts the closed-loop performance, one can let safe RL adapt $\vect{\hat S}_+$. However, this requires one to enforce~\eqref{eq:sdc_sampled} explicitly in RL. 
 The safe RL problem is then formulated as
\begin{subequations}
	\label{eq:safe_rl_problem_sampled}
	\begin{align}
		\vect{\theta}^\star := \min_{\vect{\theta}}&\ \ \sum_{k=0}^n \psi(\vect{s}_{k+1},\vect{s}_k,\vect{a}_k,\vect{\theta}) \\
		\mathrm{s.t.} & \ \ \vect{s}_{k+1} \in \vect{\hat S}_+(\vect{s}_k,\vect{a}_k,\vect{\theta}), \quad \forall \ k. \label{eq:safe_rl_problem_sampled_sdc}
	\end{align}
\end{subequations}
By enforcing~\eqref{eq:safe_rl_problem_sampled_sdc}, RL is explicitly made aware that some parameter updates are unsafe and, therefore, not feasible. Provided that the SDC holds, MPC delivers a safe policy by construction. 
Though in principle the SDC has to be enforced for each sample, in Section~\ref{sec:data} we propose an approach to largely reduce the amount of constraints.

\begin{Remark}
	The RL Problem~\eqref{eq:safe_rl_problem_sampled} is typically solved using sensitivity-based methods, hence we need to differentiate the function approximator with respect to the parameter $\vect{\theta}$. 
	In our case, we need to differentiate the robust MPC problem. This will be detailed in Section~\ref{sec:differentiability}.
\end{Remark}

The proposed safe RL framework performs the following steps: (a) at every time instant MPC is solved and differentiated; the MPC input is applied to the system; state transitions are observed and data is collected to form the sample-based SDC~\eqref{eq:sdc_sampled}; (b) the RL problem~\eqref{eq:safe_rl_problem_sampled} is solved (possibly at a lower sampling rate than MPC) and parameter $\vect{\theta}$ is updated whenever possible. A scheme is displayed in Figure~\ref{fig:rl-mpc}, with reference to the section where each component is discussed.

In this section, we have established the safe RL framework; in the next sections, we will discuss (a) how to implement robust MPC, (b) how to differentiate it in order to be able to solve Problem~\eqref{eq:safe_rl_problem_sampled}, and (c) how to manage constraint~\eqref{eq:safe_rl_problem_sampled_sdc} in a data-efficient fashion. In this paper, we address (a)-(c) by relying on a linear model of~\eqref{eq:State:Transition} to enforce safety. On the one hand, this choice makes the robust constraint satisfaction tractable and not excessively demanding in terms of computations. On the other hand, any nonlinearity present in the system will be accounted for as a perturbation, therefore introducing some conservatism. We remark that nonlinear robust MPC formulations have been developed and can be deployed within the proposed algorithmic framework. The main drawbacks of these formulations are: (a) some form of conservatism cannot be avoided and (b) the computational burden typically becomes prohibitive.

\section{Robust MPC Based on Invariant Sets}
\label{sec:rmpc}

Since guaranteeing robust constraint satisfaction in the general nonlinear case is extremely difficult~\cite{Mayne2014}, in the remainder of the paper we focus on the case of an affine model. 
We describe safety constraints~\eqref{eq:SafetyConst} as the inner approximation
\begin{align}
	\label{eq:safe_constraints}
	C\vect{s}+D\vect{a}+\vect{\bar c}\leq \vect{0}.
\end{align}
We formulate a $Q$ function approximator based on the classic robust linear MPC~\cite{Mayne2014,Chisci2001}:
\begin{subequations} %
	\label{eq:robust_mpc}%
	\begin{align}%
		\hspace{-1em}
		Q_{\vect{\theta}}(\vect{s},\vect{a}):= \hspace{-1em}& \nonumber\\
		\min_{\vect{z}} \ \ & \sum_{k=0}^{N-1} \gamma^k \left (\matr{c}{\vect{x}_k\\ \vect{u}_k}^\top H \matr{c}{\vect{x}_k\\ \vect{u}_k}  + \vect{h}^\top\matr{c}{\vect{x}_k\\ \vect{u}_k}\right ) \hspace{-4em} \nonumber \\
		&\hspace{7em}+ \gamma^N \left (\vect{x}_N^\top P \vect{x}_N + \vect{p}^\top \vect{x}_N \right )\label{eq:robust_mpc_cost} \hspace{-10em}&\hspace{1em}\\ 
		\mathrm{s.t.} \ \ & \vect{x}_0 = \vect{s}, \qquad \vect{u}_0 = \vect{a}, \label{eq:robust_mpc_ic}\\
		& \vect{x}_{k+1} = A\vect{x}_k + B \vect{u}_k + \vect{b}, & \hspace{-14em} k\in\mathbb{I}_0^{N-1}, \label{eq:robust_mpc_dyn}\\
		& C\vect{x}_k + D \vect{u}_k + \vect{c}_k \leq \vect{0}, & \hspace{-14em} k\in\mathbb{I}_0^{N-1}, \label{eq:robust_mpc_pc}\\
		& G\vect{x}_N  + \vect{g} \leq \vect{0},\label{eq:robust_mpc_tc}
	\end{align}%
\end{subequations}%
where $\vect{z}:=(\vect{x}_0,\vect{u}_0,\ldots,\vect{u}_{N-1},\vect{x}_{N})$. Note that we use $\vect{x}$, $\vect{u}$ to distinguish the MPC predictions from the actual state and action realizations $\vect{s}, \vect{a}$ which define the initial constraint~\eqref{eq:robust_mpc_ic}.
The dynamic constraints~\eqref{eq:robust_mpc_dyn} assume a nominal model without any perturbation. The tube-based approach then treats the system stochasticity, model uncertainties and safety constraint \eqref{eq:safe_constraints} by performing a suitable tightening of the path constraints~\eqref{eq:robust_mpc_pc}, i.e., $\vect{c}_k \geq \vect{\bar c}$ is used. The terminal constraints~\eqref{eq:robust_mpc_tc} are introduced to guarantee that the path constraints will never be violated at all future times $k>N$. 

The value function and optimal policy are obtained as  \cite{Gros2018}
\begin{align}
	\label{eq:value_policy}
	V_{\vect{\theta}}(\vect{s}) := \min_{\vect{a}}Q_{\vect{\theta}}(\vect{s},\vect{a}), && \pi_{\vect{\theta}}(\vect{s}) := \arg\min_{\vect{a}}Q_{\vect{\theta}}(\vect{s},\vect{a}).
\end{align}
In practice $V_{\vect{\theta}}$ and $\vect{\pi}_{\vect{\theta}}$ are computed jointly by solving~\eqref{eq:robust_mpc} without enforcing the constraint $\vect{u}_0 = \vect{a}$. 
Parameter $\vect{\theta}$ to be adapted by RL may include any of the vector and matrices defining the MPC scheme~\eqref{eq:robust_mpc}, i.e., generally
\begin{align}
\label{eq:mpc_parameters}
\vect\theta = \left\{H,\vect{h}	, P, \vect{p}, A, B, \vect{b}, \vect{\bar c}, K, \vect{\theta}_{\vect{W}}\right\}.
\end{align}
Parameters $H,P$ are typically assumed positive-definite to guarantee the solvability of~\eqref{eq:robust_mpc}. Parameters $K$ and $\vect{\theta}_{\vect{W}}$ do not appear explicitly in~\eqref{eq:robust_mpc}, and are used to compute the constraint tightening and the terminal set, which will be introduced next: we will first present the computation of the constraint tightening, and then discuss the computation of the sensitivities of $V_{\vect{\theta}}$, $Q_{\vect{\theta}}$, $\vect{\pi}_{\vect{\theta}}$ with respect to~$\vect{\theta}$.

\subsection{Recursive Robust Constraint Satisfaction}
\label{sec:constr_tightening}

In order to guarantee constraint satisfaction for the real system~\eqref{eq:State:Transition} using predictions given by the nominal model~\eqref{eq:robust_mpc_dyn}, robust MPC explains the difference between predictions and actual state transitions by means of additive noise $\vect{w}\in\vect{W}_{\vect{\theta}}$, with $\vect{W}_{\vect{\theta}}$ satisfying
\begin{align}
\label{eq:modelclosure}
\vect{\hat S}_+(\vect{s},\vect a,\vect{\theta}) = A\vect{s} + B \vect{a}+ \vect{b} + \vect{W}_{\vect{\theta}} \supseteq \vect{S}_+(\vect{s},\vect{a}).
\end{align}
We remark that, by using the affine model~\eqref{eq:robust_mpc_dyn}, conservatism is introduced as any nonlinearity present in~\eqref{eq:State:Transition} will be accounted for by $\vect{w}$. Set $\vect{W}_{\vect{\theta}}$ is parametrized by parameter $\vect{\theta}_{\vect{W}}$. Common choices in robust MPC are to parametrize $\vect{W}_{\vect{\theta}}$ using ellipsoids or polytopes; in this paper we consider the latter.

In order to perform constraint tightening, i.e., the computation of $\vect{c}_k$, we rely on the approach first proposed by~\cite{Chisci2001}. We introduce the prediction error of the nominal model~\eqref{eq:robust_mpc_dyn}: 
\begin{align*}
	\vect{E}_{k+1} &= (A-BK)\vect{E}_{k} + \vect{W}_{\vect{\theta}}, & \vect{E}_{0}&=\{\vect{0}\},
\end{align*}
where set $\vect{E}_k$ predicts an outer approximation of the dispersion set around the predicted trajectory, i.e., $\vect{S}_k^{\vect{\pi}^{\mathrm{MPC}}_{\vect{\theta},k}}\subseteq \vect{x}_k+\vect{E}_k$, $k=0,\ldots,N$, where $\vect{\pi}^{\mathrm{MPC}}_{\vect{\theta},k}:=\vect{a}_k - K\vect{e}_k$, $\vect{e}_k\in\vect{E}_k$ and
\begin{align*}
	\vect{S}_{k+1}^{\vect{\pi}^{\mathrm{MPC}}_{\vect{\theta}}} = \vect{\hat S}_+(\vect{S}_{k}^{\vect{\pi}^{\mathrm{MPC}}_{\vect{\theta},k}},\vect{a}_k - K\vect{E}_k,\vect{\theta}), && \vect{S}_{0}^{\vect{\pi}^{\mathrm{MPC}}_{\vect{\theta},k}}=\{\vect{s}_k\}.
\end{align*}
 
Feedback matrix $K$ is introduced in order to model the fact that any closed-loop strategy will compensate for perturbations on the nominal model. For ease of notation, we define $$C_K:=C-DK, \quad A_K:=A-BK.$$

Robust constraint satisfaction is then obtained provided that
\begin{align}
\label{eq:constraint_e}
& C\vect{x}_k  + D \vect{u}_k+ C_K\vect{e}_k + \vect{\bar c} \leq \vect{0}, && \forall \ \vect{e}_k \in \vect{E}_k,
\end{align}
such that $\vect{c}_k$ is obtained by adding the worst-case realization of $C_K\vect{E}_k$ to $\vect{\bar{c}}$.
For each component $i$ of the path constraint at time $k$ we define
\begin{align}
\label{eq:constr_tightening}
\vect d_{i,k} := \max_{\vect{e}} \ & (C_K)_i \vect{e}&& \mathrm{s.t.}\  \vect{e}\in\vect{E}_k \nonumber \\
= \max_{\vect{w}} \ & (C_K)_i \sum_{j=0}^{k-1} (A_K)^{j}\vect{w}_j && \mathrm{s.t.}\ \vect{w}_j\in\vect{W}_{\vect{\theta}}.
\end{align}
We lump all components $\vect{d}_{i,k}$ in vector $ \vect{d}_k$.
Then, constraint satisfaction is obtained for all $\vect{w}_k \in\vect{W}_{\vect{\theta}}$ if
\begin{align*}
\vect{c}_k = \vect{\bar c} + \vect{d}_k.
\end{align*}
If $\vect{W}_{\vect{\theta}}$ is a polytope, then~\eqref{eq:constr_tightening} can be formulated as an LP; this implies that: (a) constraint tightening is relatively cheap to compute; (b) as detailed in Section~\ref{sec:data}, the SDC enforcement becomes easier to derive.

In order to guarantee that Problem~\eqref{eq:robust_mpc} remains feasible at all times for all $\vect{w}_k \in\vect{W}_{\vect{\theta}}$, one needs to impose ad-hoc terminal conditions. More specifically, the terminal set $\mathcal{X}_{\mathrm{f}}:=\{ \, \vect{x} \, | \, G\vect{x}+\vect{g}\leq 0 \, \}$ should be robustly invariant and output admissible, i.e., there exists a terminal control law $ \vect{\kappa}_\mathrm{f}$ such that $\vect{s}_{k+1}\in\mathcal{X}_{\mathrm{f}}$ and $C\vect{s}_k+D\vect{\kappa_\mathrm{f}}(\vect{s}_k)+\vect{\bar c}\leq \vect{0}$ for every  $\vect{s}_k\in\mathcal{X}_{\mathrm{f}}$.
We consider a linear control law $\vect{\kappa}_\mathrm{f}(\vect{s}) = -K\vect{s}$, coinciding with the linear feedback used to stabilize the prediction error~$e$.

In order to compute set $\mathcal{X}_{\mathrm{f}}$, we define
\begin{align*}
\mathcal{X}_0 &:= \{ \ \vect{x} \ | \  C_K\vect{x} + \vect{c}_0 \leq 0 \ \}, \\
\mathcal{X}_k &:= \{ \ \vect{x} \in A_K \mathcal{X}_{k-1} \oplus \vect{W}_{\vect{\theta}} \ | \  C_K\vect{x} + \vect{c}_k \leq 0 \ \}.
\end{align*}
Note that, by~\eqref{eq:constraint_e}-\eqref{eq:constr_tightening}, $\vect{x}_0\in\mathcal{X}_k$ implies
\begin{align*}
	C_K(\vect{x}_j + \vect{e}_j) + \vect{\bar c} \leq 0, && \forall \, \vect{e}_j\in E_j, && \forall \, i\in\mathbb{I}_0^k.
\end{align*}
Set $\mathcal{X}_\infty$ is Maximal Robust Positive Invariant (MRPI)~\cite{Kolmanovsky1998}:
\begin{equation*}
	\vect{s}_k\in\mathcal{X}_\infty \ \ \Rightarrow \ \ A_K^{j-k}\vect{s}_{k} \in\mathcal{X}_\infty, \ \ C_K\vect{s}_j  + \vect{\bar c} \leq 0, \ \ \forall \, j>k.
\end{equation*}
Additionally, whenever the system is stable and the origin is in the interior of the constraint set, the MRPI set is finitely determined~\cite[Theorem 6.3]{Kolmanovsky1998}, i.e., $\exists \ k^\prime<\infty$ s.t. $\mathcal{X}_{k^\prime}\equiv \mathcal{X}_{k^\prime+i}$, for all $i \geq 1$. The stability requirement further motivates the introduction of feedback through matrix $K$.

As proven in~\cite{Kolmanovsky1998}, for $\vect{W}_{\vect{\theta}}$ polyhedral, $G$, $\vect{\bar g}$ are given by
\begin{align}
	\label{eq:tc_matr_vect}
	G := \matr{c}{C_K \\ C_K A_K \\ \vdots \\ C_K A_K^{k^\prime}}, && \vect{\bar g} := \matr{c}{ \vect{c}_0 \\ \vect{c}_1 \\ \vdots \\ \vect{c}_{k^\prime}}, 
\end{align}
where we stress that $\vect{c}_k = \vect{\bar c} + \vect{d}_k$, such that, by using $\vect{\kappa}_\mathrm{f} = -K\vect{e}$ one can spare a large amount of computations, $\vect{\bar g}$ being already computed. The condition $G\vect{x}_N+\vect{\bar g}\leq 0$ would then guarantee robust constraint satisfaction for all future times if $\vect{s}_N = \vect{x}_N$. However, $\vect{s}_N = \vect{x}_N+\vect{e}_N$, such that also the terminal constraint \eqref{eq:robust_mpc_tc} must be tightened. 
Analogously to the case of path constraints, we define $\vect{g} = \vect{\bar g} + \vect{h}$ with
\begin{align}
\label{eq:terminal_constr_tightening}
\vect h_{i,k} := 
\max_{\vect{w}} \ & G_i \sum_{j=0}^{k^\prime-1} A_K^{j}\vect{w}_j && \mathrm{s.t.}\ \vect{w}_j\in\vect{W}_{\vect{\theta}}.
\end{align}

\begin{Remark}
	Typically, constraints that can never become active are removed from~\eqref{eq:tc_matr_vect}, so as to reduce the dimension.
\end{Remark}

Safety is guaranteed by the following result on tube MPC.
\begin{Proposition}[Recursive Feasibility]
	\label{prop:feas_stab}
	Assume that $\mathcal{X}_\mathrm{f}:=\{ \, \vect{x} \, | \,  G\vect{x} + \vect{g} \leq 0 \, \}$ is RPI and Problem~\eqref{eq:robust_mpc} is feasible at time $k=0$. Then Problem~\eqref{eq:robust_mpc} is feasible for all $\vect{w}_k \in\vect{W}_{\vect{\theta}}$ and all times $k\geq0$. If moreover~\eqref{eq:modelclosure} holds, the real system~\eqref{eq:State:Transition} satisfies the safety constraint~\eqref{eq:safe_constraints} at all times $k\geq0$. 
\end{Proposition}
\begin{IEEEproof}
	The proof can be found in, e.g.,~\cite{Chisci2001}.
\end{IEEEproof}
This proof is valid in case $\vect{\theta}$ is kept fixed. A discussion on how to enforce recursive feasibility upon updates of $\vect{\theta}$ is proposed in Section~\ref{sec:rl}, Remark~\ref{rem:rec_feas}.

While the framework of robust linear MPC based on MRPI sets is well established, the computation of the parametric sensitivities of an MPC problem required to deploy most RL methods is not common. 
In particular, in the case of a tube-based formulation, also the constraint tightening procedure needs to be differentiated. This is also not common and deserves to be discussed in detail. 
We therefore devote the next subsection to the computation of the derivative of the value and action-value function with respect to parameter $\vect{\theta}$.

\subsection{Differentiability}
\label{sec:differentiability}
In order to be able to deploy  RL algorithms to adapt parameter $\vect{\theta}$, we need to be able to differentiate the MPC scheme and, therefore, the constraint definition with respect to $\vect{\theta}$. In principle, $\vect{\theta}$ could include $A,B,K$, but also all other parameters of the MPC formulation~\eqref{eq:mpc_parameters}. In order to compute $\nabla_{\vect{\theta}} \vect{c}_k$, $\nabla_{\vect{\theta}} C_N$, $\nabla_{\vect{\theta}} \vect{c}_N$ one can use results from parametric optimization to obtain the following lemmas~\cite{Buskens2001}. 

Consider a parametric NLP with cost $\phi_{\vect{\theta}}$, primal-dual variable $\vect{y}$ and parameter $\vect{\theta}$. We refer to~\cite{Nocedal2006} for the definition of Lagrangian $l_{\vect{\theta}}^0(\vect{y})$, KKT conditions, Linear Independence Constraint Qualification (LICQ), Second-Order Sufficient Conditions (SOSC) and Strict Complementarity (SC). For a fixed active set the KKT conditions reduce to the equality $\vect{r}^0_{\vect{\theta}}(\vect{y}) =0.$

\begin{Lemma}
	\label{lem:sensitivities}
	Consider a parametric optimization problem with optimal primal-dual solution $\vect{y}^\star$.
	Assume that LICQ, SOSC and SC hold at $\vect{y}^\star$. 
	Then, the following holds%
	\begin{align*}
	\nabla_{\vect{\theta}} \phi_{\vect{\theta}} = \nabla_{\vect{\theta}} l^0_{\vect{\theta}}(\vect{y}), && \frac{\partial \vect{r}^0_{\vect{\theta}}}{\partial \vect{y}} \frac{\mathrm{d}}{\mathrm{d}\vect{\theta}}\vect{y}^\star = \frac{\partial \vect{r}^0_{\vect{\theta}}}{\partial \vect{\theta}}.
	\end{align*}
\end{Lemma}
\begin{IEEEproof}
	The result can be found in, e.g.,~\cite{Buskens2001}.
\end{IEEEproof}

\begin{Corollary}
	\label{cor:sensitivities}
	Assume that LICQ, SOSC and SC hold at the optimal solution of~\eqref{eq:robust_mpc}. 
	Then, the value function $V_{\vect{\theta}}$, action-value function $Q_{\vect{\theta}}$ and optimal solution $\vect{y}^\star$ (therefore also policy $\vect\pi$) are differentiable with respect to parameter ${\vect{\theta}}$, with:
	\begin{subequations}
	\label{eq:sensimilla}
	\begin{align}
	&\nabla_{\vect{\theta}} V_{\vect{\theta}}(\vect{s}) = \nabla_{\vect{\theta}} \bar l_{\vect{\theta}}(\vect{y}), && 		\nabla_{\vect{\theta}} Q_{\vect{\theta}}(\vect{s},\vect{a}) = \nabla_{\vect{\theta}} l_{\vect{\theta}}(\vect{y}), \label{eq:sensQV} \\
	&\frac{\partial \vect{r}_{\vect{\theta}}}{\partial \vect{y}} \frac{\mathrm{d}}{\mathrm{d}\vect{\theta}}\vect{y}^\star = \frac{\partial \vect{r}_{\vect{\theta}}}{\partial \vect{\theta}},
	\end{align}
	\end{subequations}
	where $l_{\vect{\theta}}$ is the Lagrangian of Problem~\eqref{eq:robust_mpc}, $\bar  l_{\vect{\theta}}$ is the Lagrangian when constraint $\vect{u}_0=\vect{a}$ is eliminated, and $\vect{r}_{\vect{\theta}}$ denotes the KKT conditions for the optimal active set.
\end{Corollary}

\begin{Remark}
	When solving an LP, QP or NLP using a second-order method, e.g., active-set or interior-point, the most expensive operation is the factorization of the KKT matrix $\frac{\partial \vect{r}_{\vect{\theta}}}{\partial \vect{y}}$. Once the matrix is factorized, the solution of the linear system is computationally inexpensive. Therefore, the sensitivities of the solution are in general much cheaper to evaluate than solving the problem itself. The sensitivity of the optimal value function is even simpler to compute, since it consists in the differentiation of the Lagrangian, see~\eqref{eq:sensQV}. 
\end{Remark}

In~\eqref{eq:sensimilla}, $\vect{r}_{\vect{\theta}}$, $l_{\vect{\theta}}$, and $\bar l_{\vect{\theta}}$ depend on $\vect{c}_k$, $\vect{g}$ which, in turn, depend on $\vect{\theta}$ as they are optimal values of parametric optimization problems~\eqref{eq:constr_tightening} and~\eqref{eq:terminal_constr_tightening}. Consequently, one needs to evaluate $\nabla_{\vect{\theta}} \vect{c}_k$, $\nabla_{\vect{\theta}} \vect{g}$. In the following, we further detail the application of Lemma~\ref{lem:sensitivities} to this case.

We consider only Problem~\eqref{eq:constr_tightening}, since the derivation for~\eqref{eq:terminal_constr_tightening} is analogous. 
First, we state separability and, therefore, parallelizability of the computation of $\vect{d}_{k}$ in the following Lemma. 
\begin{Lemma}
	Each component of $\vect{d}_{k}$ can be computed as
	\begin{align}
	\vect d_{i,k} &= \sum_{j=0}^{k-1} \vect d_{i,k\change{,j}}, \text{ where} \nonumber \\
	\label{eq:constr_tightening_subproblem}
	\vect d_{i,k\change{,j}} & := 
	\max_{\vect{w}_j} \  (C_K)_i A_K^j \vect{w}_j \qquad \mathrm{s.t.}\qquad \vect{w}_j\in\mathcal{W}_{\vect{\theta}}.
	\end{align}
\end{Lemma}
\begin{IEEEproof}
	Each term in the sum $\sum_{j=0}^{k-1} A_K^{j}\vect{w}_j $ depends only on variable $\vect{w}_j$, and the problem is fully separable.
\end{IEEEproof}
Then, Lemma~\ref{lem:sensitivities} can be applied to obtain
\begin{align*}
\frac{\mathrm{d} \vect{d}_{k}}{\mathrm{d} \vect{\theta}} &= \left (\frac{\mathrm{d} \vect d_{1,k}}{\mathrm{d} \vect{\theta}},\ldots,\frac{\mathrm{d} \vect d_{n_{\vect{c}_k},k}}{\mathrm{d} \vect{\theta}}\right ),\\
\frac{\mathrm{d} \vect d_{i,k}}{\mathrm{d} \vect{\theta}} &= \sum_{j=0}^{k-1} \frac{\mathrm{d} \vect d_{i,k,j}}{\mathrm{d} \vect{\theta}}, \qquad \frac{\mathrm{d} \vect d_{i,k,j}}{\mathrm{d} \vect{\theta}} = \frac{\mathrm{d} l^{\vect{d}}_{\vect{\theta},i,j}}{\mathrm{d} \vect{\theta}} ,
\end{align*}
where 
$l^{\vect{d}}_{\vect{\theta},i,j}$ is the Lagrangian of Problem~\eqref{eq:constr_tightening_subproblem}. 
Provided that a second-order method is used for solving Problem~\eqref{eq:constr_tightening_subproblem}, then the matrix factorization is available and can be reused to compute the sensitivities at a negligible cost.
For any function $\vect{f}(\vect{c}_k(\vect{\theta}))$, the chain rule yields
\begin{align*}
\frac{\mathrm{d} \vect{f}}{\mathrm{d}  \vect{\theta}} = \frac{\mathrm{d} \vect{f}}{\mathrm{d} \vect{c}_k} \frac{\mathrm{d} \vect{c}_k }{\mathrm{d} \vect{\theta}}= \frac{\mathrm{d} \vect{f}}{\mathrm{d} \vect{c}_k}\frac{\mathrm{d} \vect{d}_{k} }{\mathrm{d} \vect{\theta}}.
\end{align*}

\begin{Remark}
	As underlined above, Problems~\eqref{eq:constr_tightening_subproblem} can be solved in parallel not only for each prediction time $k$, but also for each component $i$. Moreover, if an active-set solver is used, the active set can be initialized and the matrix factorization reused, such that often there will be no need for recomputing the factorization and computations can be done in an extremely efficient manner. Finally, Problems~\eqref{eq:constr_tightening_subproblem} are very low dimensional and are therefore solved extremely quickly.
\end{Remark}

\subsection{Guaranteeing MPC Feasibility and LICQ}
\label{sec:slack}

We detail next two issues that can be easily encountered when deploying RL based on MPC~\eqref{eq:robust_mpc}. Since these are common, a simple solution that has become a standard in MPC is readily available.

\paragraph{MPC Feasibility} Since the set of possible perturbations is not known a priori but rather approximated as $\vect{W}_{\vect{\theta}}$ based on the collected samples, it cannot be excluded that some future sample $\vect w_k$ will not be inside the set, i.e., $\vect{w}_k\notin \vect{W}_{\vect{\theta}}$. The set approximation must then be modified to include the new sample (see Section~\ref{sec:data_compression} and~\eqref{eq:sdc_adaptation}), but recursive feasibility is potentially lost, no action is computed and the controller stops working. 

\paragraph{Sensitivity Computation} The sensitivity computation is valid only if LICQ holds. However, Problem~\eqref{eq:robust_mpc} is not guaranteed to satisfy LICQ.

We propose to address both issues by using a common approach in MPC, i.e., a constraint relaxation for Constraints~\eqref{eq:robust_mpc_pc} and~\eqref{eq:robust_mpc_tc} with an exact penalty~\cite{Scokaert1999a}: variables $\vect{\sigma}_k$ are introduced, the constraints are modified as
\begin{subequations}
	\label{eq:pc_relaxed}
	\begin{align}
		C\vect{x}_k + D\vect{u}_k + \vect{c}_k \leq \vect{\sigma}_k,  \\
		G\vect{x}_N + \vect{g} \leq \vect{\sigma}_N, && \vect{\sigma}_k \geq 0, &&k\in\mathbb{I}_1^{N},
	\end{align}
\end{subequations}
and the term $\sum_{k=1}^N \vect{\rho}^\top \vect{\sigma}_k$ is added to the cost. 
\begin{Remark}
	Constraints only involving the controls do not need to be relaxed, while any constraint involving the states should not be imposed at $k=0$, since then LICQ can not be guaranteed even if the relaxation proposed above is deployed. 
\end{Remark}

We formalize the result in the next proposition.
\begin{Proposition}
	\label{prop:feas_licq}
	Assume that Constraints~\eqref{eq:robust_mpc_pc} and~\eqref{eq:robust_mpc_tc} are relaxed as per~\eqref{eq:pc_relaxed} and the term $\sum_{k=1}^N \vect{\rho}^\top \vect{\sigma}_k$ is added to the cost. Then, if $\vect{\rho}<\infty$ is large enough, the solution is unchanged whenever feasible and recursive feasibility and LICQ are guaranteed.
\end{Proposition}
\begin{IEEEproof}
	The first result, i.e., solution equivalence whenever feasible and recursive feasibility is well-known~\cite{Scokaert1999a} and~\cite[Theorem 14.3.1]{Fletcher1987}. Regarding LICQ, we first note that in a formulation without Constraints~\eqref{eq:robust_mpc_pc} and~\eqref{eq:robust_mpc_tc} LICQ holds by construction: Constraints~\eqref{eq:robust_mpc_ic} and~\eqref{eq:robust_mpc_dyn} can be eliminated by condensing~\cite{Bock1983}, yielding a problem with $Nn_{\vect{u}}$ unconstrained variables. The introduction of any linearly independent set of pure control constraints does then not jeopardize LICQ by construction. 
	Assume now to introduce a linearly-dependent constraint of the form~\eqref{eq:robust_mpc_pc} or~\eqref{eq:robust_mpc_tc} with Jacobian $\vect{\nu}$. By introducing slack variable $\sigma$, the new Jacobian becomes $\matr{cc}{\vect{\nu} & 1}$, which is by construction linearly independent with $\matr{cc}{\vect{\nu} & 0}$ and, consequently, with the other constraints in the problem. 
\end{IEEEproof}
As discussed in Section~\ref{sec:safe_rl}, safety holds with probability $\mathbb{P}\left [ \, \eqref{eq:sdc} \, | \, \vect{\theta}  \, \right ] =1$. This is an intrinsic  issue of any safety-enforcing control scheme, and the proposed relaxation only solves the issue of avoiding infeasibility for the MPC scheme. This is further discussed in Section~\ref{sec:discussion}.

\section{Safe Design Constraint and Data Management}
\label{sec:data}

As explained in the previous section, safety is obtained if all possible state transitions are correctly captured in the MPC formulation, i.e., if $\vect{W}_{\vect{\theta}}$ and, therefore, $\vect{g}_{\vect{\theta}}$ is correctly identified. 
In other words, based on the collected data the SDC must be enforced in order to guarantee that the uncertainty described by $\vect{W}_{\vect{\theta}}$ is representative of the real system~\eqref{eq:State:Transition}. In this section we propose a sample-based formulation of the SDC~\eqref{eq:sdc} to be used within RL formulations~\eqref{eq:rl_problem_sampled}.

Many control systems are typically operated at high sampling rates and, consequently, data is collected at high rates. In order to be able to deal in real-time with the large amounts of data that cumulate, it is necessary to (a) retain only strictly relevant data, and (b) compress the available information using appropriately defined data structures. In the following, we first discuss the data structures involved in RL-MPC and then discuss how to make an efficient use of data in the context of the proposed MPC formulation.

\subsection{Set Membership and SDC}

Consider the (possibly very large) set of state transitions observed on the real system:
\begin{align}
	\mathcal{D} = \left\{\left(\vect{s}_1,\vect{a}_1, \vect{s}_2\right),\ldots, \left(\vect{s}_{n},\vect{a}_n, \vect{s}_{n+1}\right)\right\}.
\end{align}
The problem of enforcing the SDC~\eqref{eq:sdc_sampled} is related to the one of estimating the dispersion set $\vect{S}_+$ from data, which has been studied in the context of \emph{set-membership system identification} (SMSI)~\cite{Bertsekas1971a}. 
Essentially, the dispersion set must satisfy
\begin{align}
	\label{eq:set_membership}
	\vect{s}_+ \in \vect{\hat S}_+(\vect{s},\vect{a},\vect{\theta}), && \forall \ \left(\vect{s},\vect{a}, \vect{s}_+\right) \in \mathcal{D},
\end{align}
i.e., $\vect{\theta}$ must satisfy the SDC~\eqref{eq:sdc_sampled}, and therefore belongs to set
\begin{align}
\label{eq:sdc_definition_param}
\mathcal{S}_\mathcal{D} := \left\{ \vect{\theta} \,|\, \vect{g}_{\vect{\theta}}(\vect{s}_{k+1},\vect{s}_k,\vect{u}_k) \leq 0,
\, \forall\, (\vect{s}_+,\vect{s},\vect{u})\in \mathcal{D} \right\}.
\end{align}

We should underline here the difference between $\vect{\hat S}_+$ defined in~\eqref{eq:param_disp_set} and $\mathcal{S}_\mathcal{D}$. The former is an outer approximation of the set of all realizations $\vect{s}_+$, given state and action $\vect{s},\vect{a}$, parametrized by $\vect{\theta}$. The latter instead describes the set of parameters $\vect{\theta}$ such that all state transitions from $\mathcal{D}$ are contained in $\vect{\hat S}_+$. 

In SMSI parameter $\vect{\theta}$ is typically selected so as to obtain the smallest possible set $\vect{S}_+$. In the absence of specific information on the control task to be executed this is arguably a very reasonable approach. However, given a specific control task to be executed, better performance might be obtained by selecting parameter $\vect{\theta}$ to approximate some part of $\vect{S}_+$ accurately even at the cost of increasing the volume of $\vect{\hat S}_+$. We therefore provide the following definition.
\begin{Definition}[Set Membership Optimality]
	\label{def:sdc_optimality}
	Equivalently to~\eqref{eq:rl_problem}, we define the closed-loop cost $J(\vect\pi_{\vect{\theta}})$ associated with the policy $\vect\pi_{\vect{\theta}}$ learned by RL when relying on the set $\vect{\hat S}_+(\vect{s},\vect{a},\vect{\theta})$.
	Then, parameter $\vect{\theta}$ is (locally) optimal for the control task iff $\nexists \ \vect{\bar \theta} \ \in \mathcal{B}_\epsilon(\vect{\theta}) \ \mathrm{s.t.} \ J(\vect\pi_{\vect{\bar \theta}}) < J(\vect\pi_{\vect{\theta}})$, where $\mathcal{B}_\epsilon(\cdot)$ denotes a ball of radius $\epsilon$ centered at $\vect\theta$. 
\end{Definition}

In other words, the optimal parameter minimizes the cost subject to the SDC~\eqref{eq:sdc_sampled}. 
Based on this definition, the optimal set approximation is obtained by an SMSI which selects $\vect{\theta}$ to maximize the closed-loop performance of the policy $\vect\pi_{\vect\theta}$. In this paper, we aim at doing so by means of RL. 

Every time the RL problem~\eqref{eq:safe_rl_problem_sampled} is solved, one must ensure that $\vect{\theta}\in \mathcal{S}_\mathcal{D}$, i.e., $|\mathcal{D}|$ constraints need to be included in the problem formulation. With large amounts of data, this could make the problem computationally intractable. In the following, we will analyze how to tackle this issue.

While the previous developments did not require any assumption on function $\vect{g}_{\vect{\theta}}$, in order to efficiently manage data we will assume hereafter that $\vect{g}_{\vect{\theta}}$ is affine in $\vect{s}, \vect{a}$. Note that this choice is consistent with the choice of an affine model. The SDC then becomes
\begin{align*}
\mathcal{S}_\mathcal{D} &:= \left\{ \vect{\theta} \,|\,  M \left (\vect{s}_{+} - A\vect{s} - B\vect{a} - \vect{b} \right ) \leq  \vect{m}, \, \forall \, (\vect{s}_+,\vect{s},\vect{u})\in \mathcal{D} \right\},
\end{align*}
for some $M,\vect{m}$ possibly part of $\vect\theta$. Since both $\mathcal{S}_\mathcal{D}$ and $\vect{W}_{\vect{\theta}}$ are defined based on  $\vect{g}_{\vect{\theta}}$, parameters $M,\vect{m}$ for the two sets coincide. Therefore, the SDC directly defines the uncertainty set $\vect{W}_{\vect{\theta}}$ used by MPC to compute safe policies. Note that $A$, $B$, $\vect{b}$, $M$, $\vect{m}$ can all be included in $\vect{\theta}$ and, therefore, adapted by RL. Which parameter to adapt or keep constant is a design choice, see also Remark~\ref{rem:model_adaptation}.

\subsection{Model-Based Data Compression}
\label{sec:data_compression}

In this section we discuss how to compress the available data without loss of information to significantly reduce the complexity of safe RL. It will become clear that the nominal model~\eqref{eq:robust_mpc_dyn} plays a very important role in this context, as it makes it possible to organize data such that it can be efficiently exploited. Unfortunately, this efficiency is lost if the model parameters are updated. Approaches to circumvent this issue can be devised and are the subject of ongoing research. 
We begin the analysis by providing the following definition. 
\begin{Definition}[Optimal Data Compression]
	\label{def:data_compression}
	Given the selected parametrization and MPC formulation, an \emph{optimal data compression} selects a dataset $\mathcal{\bar D}\subseteq\mathcal{D}$ such that:
		\begin{align}
			\label{eq:data_efficiency}
			\big |\mathcal{\bar D}\big | = \min_{\mathcal{\hat D}} \ \ & \big |\mathcal{\hat D}\big | &
			\mathrm{s.t.} \ \ & \mathcal{S}_\mathcal{\hat D}\equiv \mathcal{S}_\mathcal{D}.
		\end{align}
\end{Definition}
Hence an optimal data compression retains the minimum amount of data required to represent the set $\mathcal{S}_{\mathcal D}$. 
In the following, we assume that $A$, $B$, $\vect{b}$ are fixed and exploit the model to achieve an optimal data compression.

The introduction of the nominal model~\eqref{eq:robust_mpc_dyn} allows us to restructure the data and only store the noise $\mathcal{W}:=\{ \, \vect{w}_0,\ldots, \vect{w}_n \,\}$, obtained from 
\begin{align}
	\label{eq:noise_samples}
	\vect{w}_k = \vect{s}_{k+1} - \left (A\vect{s}_k + B \vect{a}_k + \vect{b}\right ).
\end{align}
By using~\eqref{eq:noise_samples}, we rewrite the SDC as
\begin{align}
	\mathcal{S}_\mathcal{D} = \mathcal{S}_\mathcal{W} &:= \left\{ \vect{\theta} \,|\,  M \vect{w} \leq  \vect{m}, \, \forall \, \vect{w}\in \mathcal{W} \right\},
	\label{eq:sdc_polytope}
\end{align}
with $\vect{\theta}_{\vect{W}}=(M, \vect{m})$ a component of $\vect{\theta}$. 
By exploiting the model we can therefore reduce the dimension of the space of the dataset from $2n_{\vect{s}} + n_{\vect{a}}$ to $n_{\vect{s}}$, since $(\vect{s},\vect{a},\vect{s}_+) \in \mathbb{R}^{n_{\vect{s}} \times n_{\vect{a}} \times n_{\vect{s}}}$ and $\vect{w} \in \mathbb{R}^{n_{\vect{s}}}$. 
Note that this is only possible if one neglects the dependence of $\vect{W}_{\vect{\theta}}$ (and therefore $\mathcal{W}$) on $\vect{s},\vect{a}$. 

The dispersion set approximation related with~\eqref{eq:sdc_polytope} is then
\begin{align}
	\vect{\hat S}_+(\vect{s},\vect{a},\vect{\theta}) 
	&= \{ \, A\vect{s} + B\vect{a} + \vect{b} + \vect{w} \,|\, \forall \, \vect{w} \, \mathrm{s.t.} \, M \vect{w} \leq  \vect{m} \, \},
	\label{eq:disp_set_affine}
\end{align}
such that~\eqref{eq:constr_tightening}, \eqref{eq:terminal_constr_tightening} used in constraint tightening are LPs. 
Additionally, it is cheap to (a) evaluate if $\vect{s}_+\in\vect{\hat S}_+$ by using \eqref{eq:noise_samples} and (b) verifying that $\vect{\theta} \in \mathcal{S}_\mathcal{W}$, i.e., $M \vect{w}_k \leq  \vect{m}$ holds, since both operations only require few matrix-vector operations. However, while (a) requires a single evaluation of the inequality in~\eqref{eq:disp_set_affine}, (b) requires to evaluate the inequality for each data point in~\eqref{eq:sdc_polytope}. 

The use of the model and the convexity assumption make it possible to further compress data: any point in the interior of the convex hull of set $\mathcal{W}$ does not provide any additional information regarding~\eqref{eq:sdc_polytope}, such that the convex hull
\begin{align*}
	\mathcal{\bar W} := \mathrm{Conv}(\mathcal{W}),
\end{align*}
carries all necessary information. 
Constructing the convex hull facet representation can be a rather expensive operation, which can in general not be done online. 
However, checking whether a sample $\vect{w}_k$ lies inside the convex hull $\mathcal{\bar W}$ of a set of samples $\mathcal{W}$ can be done without building the facet representation of the convex hull. 
To this end, we define the LP
\begin{align}
	\label{eq:chull_lp}
	\zeta := \min_{\vect{z}} \ \ & \sum_{i=1}^{|\mathcal{\bar W}|} \vect{z}_i &
	\mathrm{s.t.} \ \ & \vect{\hat w} = \sum_{i=1}^{|\mathcal{\bar W}|} \vect{z}_i \vect{w}_i, && \vect{z} \geq 0,
\end{align}
and exploit the following result. 
\begin{Proposition}
	Assume that $\vect{0}\in \mathrm{Conv}(\mathcal{W})$, then
	$$\zeta \leq 1 \Leftrightarrow \vect{\hat w} \in \mathrm{Conv}(\mathcal{W}),$$
	with $\zeta$ from~\eqref{eq:chull_lp}. 
\end{Proposition}
\begin{IEEEproof}
	The definition of convex hull implies $\vect{\hat w} \in \mathrm{Conv}(\mathcal{W}) \Rightarrow \exists \ \vect{z} \geq0, \|\vect{z}\|_1 = 1 \ \mathrm{s.t.} \ \vect{\hat w} = \sum_{i=1}^{|\mathcal{\bar W}|} z_i \vect{w}_i$, which proves $\zeta \leq 1 \Leftarrow \vect{\hat w} \in \mathrm{Conv}(\mathcal{W})$.
	This also covers the implication $\zeta = 1 \Rightarrow \vect{\hat w} \in \mathrm{Conv}(\mathcal{W})$.
	The implication $\zeta < 1 \Rightarrow \vect{\hat w} \in \mathrm{Conv}(\mathcal{W})$ is proven by noting that $\zeta \vect{w}_i \in \mathrm{Conv}(\mathcal{W}), \ \forall \ \zeta \in [0,1]$. The case $\zeta < 1$ is thus immediately reconducted to the case $\zeta = 1$.
\end{IEEEproof}
Formulation~\eqref{eq:chull_lp} is very convenient, as it is always feasible, provided that $\mathcal{W}$ spans the full space $\mathbb{R}^{n_{\vect{w}}}$, which is a minimum reasonable requirement in this context, since it guarantees that every vector in $\mathbb{R}^{n_{\vect{w}}}$ can be obtained as a linear combination of $\vect{w}_i$. Note also that the assumption $\vect{0}\in \mathrm{Conv}(\mathcal{W})$ is a rather mild requirement on the accuracy of the model, as it always holds when standard system identification techniques are deployed to estimate the model parameters $A,B,\vect{b}$. 

We prove the efficiency of the convex hull approach in the following theorem.
\begin{theorem}[Convex Hull Optimality]
	\label{thm:efficiency}
	Given the dispersion set $\vect{\hat S}_+$ defined in~\eqref{eq:disp_set_affine}, the convex hull $\mathcal{\bar W}$ of the state transition noise $\mathcal{W}$ is an optimal data compression.
\end{theorem}
\begin{IEEEproof}
	The definition of $\mathcal{\bar W}$ implies that any point in $\mathcal{W}$ can be obtained as the convex combination of points in $\mathcal{\bar W}$. By definition we have that $\vect{g}_{\vect{\theta}}^{\vect{w}}(\vect{w}_1)\leq 0$, $\vect{g}_{\vect{\theta}}^{\vect{w}}(\vect{w}_2)\leq 0$, for all $\vect{w}_1,\vect{w}_2 \in \mathcal{W}$. For $\beta\in [0,1]$ we define $\vect{w}_\beta:=\beta \vect{w}_1 + (1-\beta)\vect{w}_2$ such that
	\begin{align*}
		\vect{g}_{\vect{\theta}}^{\vect{w}}(\vect{w}_\beta) \leq \beta \vect{g}_{\vect{\theta}}^{\vect{w}}(\vect{w}_1) + (1-\beta) \vect{g}_{\vect{\theta}}^{\vect{w}}(\vect{w}_2) \leq 0,
	\end{align*}
	since convexity of $\mathcal{\bar W}$ is equivalent to convexity of $\vect{g}_{\vect{\theta}}^{\vect{w}}$. This entails that $\vect{w}_\beta\in\mathcal{W}$, such that any set $\vect{\hat S}_+$ computed using $\mathcal{\bar W}$ satisfies 
	\begin{align*}
		\vect{s}_{k+1} \in \vect{\hat S}_+(\vect{s}_k,\vect{u}_k,\vect{\theta}),&&  \forall\, (\vect{s}_{k+1},\vect{s}_k,\vect{u}_k)\in \mathcal{D},
	\end{align*}
	i.e., $ \forall\, \vect{w}\in \mathcal{W}$. 
	Finally, by removing any data point from $\mathcal{\bar W}$ one cannot guarantee that $\mathcal{S}_\mathcal{\bar W}=\mathcal{S}_\mathcal{W}$, such that $\mathcal{\bar W}$ solves~\eqref{eq:data_efficiency}. 
\end{IEEEproof}

In order to construct the convex hull $\mathcal{\bar W}$, once a new sample $\vect{w_k}$ is available, we check whether $\vect{w}_k\in\mathcal{\bar W}$ by solving~\eqref{eq:chull_lp}. In case $\zeta \leq 1$, $\mathcal{\bar W}$ does not need to be updated; otherwise, the new point is added to $\mathcal{W}$, such that $\mathcal{\bar W}$ is implicitly updated. Note that the LP~\eqref{eq:chull_lp} has linear complexity in the amount of vertices $|\mathcal{\bar W}|$. 

\begin{Remark}
	\label{rem:sdc_dimension}
	As data are collected, $|\mathcal{D}|$ becomes indefinitely large.
	Even though $\mathcal{\bar W}$ is optimal, also $|\mathcal{\bar W}|$ can become indefinitely large, though arguably at a lower rate than $|\mathcal{D}|$. This is a fundamental issue of any sample-based SDC or set membership approach, unless additional assumptions are introduced. Simple strategies such as limiting the maximum amount of vertices/facets to be used for an approximation $\mathcal{\hat W}$ of the convex hull could be devised, at the price of introducing conservatism. Such investigations are beyond the scope of this paper and require future research.
\end{Remark}

\begin{Remark}
	\label{rem:model_adaptation}
	One must take extra care if the model parameters $A,B$ are updated. Indeed, if one updates $A,B,\vect{b}$ to $\tilde A,\tilde B,\vect{\tilde b}$, then the new noise satisfies:
	\begin{align*}
		\vect{\tilde w}_{k} &= \vect{s}_{k+1} - \tilde A\vect{s}_{k} - \tilde B\vect{a}_{k} - \vect{\tilde b},
	\end{align*}
	such that the noise update
	\begin{align*}
	\vect{\tilde w}_{k} - \vect{w}_{k} &= ((A-\tilde A)\vect{s}_{k} + (B - \tilde B)\vect{a}_{k} + \vect{b} - \vect{\tilde b})
	\end{align*}
	is state-action dependent for all $\tilde A\neq A$, $\tilde B\neq B$.
	Therefore, any change in those parameters requires one to recompute the noise vector $\vect{w}_k$ for all recorded state-action pairs. 
	Updates in parameter $\vect{b}$ instead can be performed without much complication and simply entail a shift of the noise, which is state-action independent, such that $\mathcal{\tilde W} = \mathcal{W}+\vect{b} - \vect{\tilde b}$. 
\end{Remark}

\subsection{Further Observations on the Sample-Based SDC}

The convex hull $\mathcal{\bar W}$ is the smallest set encompassing all observed samples, i.e., it is the smallest set satisfying the SDC~\eqref{eq:sdc_sampled}. Therefore, it is optimal both in terms of volume and in terms of cost, i.e., in the sense of Definition~\ref{def:sdc_optimality}. Hence, one might be tempted to select $\vect{W}_{\vect{\theta}}=\mathcal{\bar W}$ to reduce conservatism in MPC~\eqref{eq:robust_mpc} as much as possible. However, $\mathcal{\bar W}$ is typically composed of a very large amount of facets, which renders the constraint tightening procedure very costly and results in a terminal constraint of high dimension. In practice, a set $\vect{W}_{\vect{\theta}}$ of fixed and low complexity is preferred, hence the importance of enforcing set membership optimality as per Definition~\ref{def:sdc_optimality}, i.e., through RL.

Thus far we have not discussed how the set $\vect{W}_{\vect{\theta}}$ is represented. However, the choice of representation becomes important when dealing with large amounts of data. 
Moreover, the parametrization of $\vect{W}_{\vect{\theta}}$ should be selected consistently with the algorithm used for solving the robust MPC Problem. Convex polytopes can be parametrized using the so-called (a) facet or (b) vertex representation.
In case (a), the set is parametrized as $\vect{W}_{\vect{\theta}}:=\{ \, \vect{w} \,|\, M\vect{w} \leq \vect{m} \, \}$ with parameter $\vect{\theta}_{\vect{W}}=(M,\vect{m})$. In case (b), the set is parametrized as $\vect{W}_{\vect{\theta}}:=\{ \, \vect{w} \,|\, \vect{w}\in \mathrm{Conv}(\{\vect{v}_0,\ldots,\vect{v}_m\}) \, \}$ with parameter $\vect{\theta}_{\vect{W}}=(\vect{v}_0,\ldots,\vect{v}_m)$, i.e., the vertices the polytope. 

Differently from $\vect{W}_{\vect{\theta}}$, for the convex hull $\mathcal{\bar W}$ we use the vertex representation. This allows a simpler and less computationally demanding construction and incremental update of $\mathcal{\bar W}$. This advantage, however, results in a more costly evaluation of $\vect{w}_k\in\mathcal{\bar W}$ with respect to a facet representation. 
Finally, this makes it simple to enforce the SDC~\eqref{eq:sdc_polytope}, which becomes
\begin{align}
	\label{eq:sdc_conv_hull}
	\mathcal{S}_\mathcal{\bar W} &= \left\{ \vect{\theta} \,|\,  M \vect{w} \leq  \vect{m}, \, \forall \, \vect{w}\in \mathcal{\bar W} \right\}.
\end{align}
The question on which representation is the most convenient for the convex hull $\mathcal{\bar W}$ is still open and will be further investigated in future research, which will also consider a combination of both the facet and vertex representation. We provide next some fundamental observations regarding $\mathcal{\bar W}$, its cardinality and its relationship with the nominal model.
\begin{Remark}
	\label{rem:violation}
	As a consequence of Fundamental Limitation~\ref{lim:safety},  in any sample-based context it is possible that a new sample $\vect{w}_k$ falls out of the convex hull of previous samples, i.e., $\vect{w}_k\notin\mathcal{\bar W}$, such that potentially $M_i \vect{w}_k \geq  \vect{m}_i$ for some $i$. In this case, one needs to instantaneously adapt the SDC~\eqref{eq:sdc_sampled}. A straightforward adaptation of $\mathcal{S}_\mathcal{\bar W}$ is obtained as
	\begin{align}
		\label{eq:sdc_adaptation}
		\vect{m}_i \leftarrow \max(\vect{m}_i,M_i \vect{w}_k).
	\end{align}
	This enlargement of the uncertainty set $\vect{W}_{\vect{\theta}}$ entails an enlargement of the dispersion set, such that the constraints will be further tightened. This can jeopardize recursive feasibility of MPC~\eqref{eq:robust_mpc} such that it is necessary to deploy the constraint relaxation proposed in Section~\ref{sec:slack}. 
\end{Remark}

\section{Safe RL MPC Implementation}
\label{sec:rl}

After having introduced all necessary components of the RL-MPC scheme, in this section we focus on the safe RL problem, discuss more in detail the RL problem and present some open research questions. 

Most recursive RL approaches use only the current sample and update the parameter recursively as
\begin{align}
	\label{eq:rl_param_update}
	\vect{\theta}_{k+1} = \vect{\theta}_k + \alpha (\vect{\theta}^*-\vect{\theta}_k),
\end{align}
with $\vect{\theta}^*=\vect{\theta}_k + \nabla_{\vect{\theta}}\psi(\vect{s}_{k+1},\vect{s}_k,\vect{a}_k,\vect{\theta})$ and $\psi$  defined, e.g., as per~\eqref{eq:q_learning} or~\eqref{eq:policy_gradient}.

This means that the update is the solution (or a step of stochastic gradient descent) of an unconstrained optimization problem. 
Since we need to enforce the SDC, the safe RL problem~\eqref{eq:safe_rl_problem_sampled} yielding $\vect{\theta}^*$ is constrained. By relying on the developments of Section~\ref{sec:data}, we formulate~\eqref{eq:safe_rl_problem_sampled} as
\begin{subequations}
	\label{eq:safe_rl_sampled}
	\begin{align}
	\hspace{-0.5em}\min_{\vect{\theta}} \ & \psi(\vect{s}_{k+1},\vect{s}_k,\vect{a}_k,\vect{\theta}),
	\label{eq:safe_rl_sampled_cost}\\
	\hspace{-0.5em}\mathrm{s.t.} \ & H \succ 0, \qquad P \succ 0, \label{eq:safe_rl_sampled_pd} \\
	&M \vect{w} \leq  \vect{m}, \qquad \forall \, \vect{w}\in \mathcal{\bar W}. \label{eq:safe_rl_sampled_sdc}
	\end{align}
\end{subequations}
Problem~\eqref{eq:safe_rl_sampled_cost} (without constraints) can be seen as a standard RL formulation~\cite{Zanon2019}. Positive-definite constraints~\eqref{eq:safe_rl_sampled_pd} are introduced to make sure that MPC is properly formulated and easily and efficiently solvable.
The SDC~\eqref{eq:sdc_sampled} is imposed in~\eqref{eq:safe_rl_sampled_sdc}. 
If the model parameters $A$, $B$ are updated, the SDC cannot be implemented as~\eqref{eq:sdc_sampled}, but rather as $\vect{\theta}\in \mathcal{S}_\mathcal{D}$, with $\mathcal{S}_\mathcal{D}$ given by~\eqref{eq:sdc_definition_param}.

\begin{Remark}
	\label{rem:rec_feas}
	Any update in parameter $\vect{\theta}$ results in a modification of set $\vect{W}_{\vect{\theta}}$, such that the existence of a solution to the MPC problem could be jeopardized. Several options can be envisioned, including: (a) updating $\vect{\theta}$ only when feasible, (b) reducing the step size until feasibility is recovered, (c) enforcing feasibility as an additional constraint in~\eqref{eq:safe_rl_sampled}. Any of the strategies (a)-(c) guarantees the feasibility of the updated robust MPC scheme since, provided that MPC is correctly formulated, by Proposition~\ref{prop:feas_stab} initial feasibility entails recursive feasibility. In turn, this entails safety of the parameter update. It is worth mentioning that we did not encounter feasibility issues in the simulations we performed. Nevertheless, future research will investigate this issue in depth. 
\end{Remark}

\begin{Remark}
	Both $Q_{\vect{\theta}}$ and $J(\vect{\pi}_{\vect{\theta}})$ depend on the constraint tightening procedure, such that their first-order sensitivities can be discontinuous. Consequently, also the first-order sensitivities of $\psi$ are non-smooth. 
	This is the case when SC does not hold either in the MPC problem~\eqref{eq:robust_mpc} or in the constraint tightening problems~\eqref{eq:constr_tightening},~\eqref{eq:terminal_constr_tightening}, i.e., 
	when some constraint(s) are weakly active, such that infinitesimal perturbations could cause an active-set change. 
	In principle, this could create problems to algorithms for continuous optimization. However, the set on which SC does not hold has zero measure, such that the probability that one sample falls onto one of these points is zero, and the RL solution is unaffected. 
\end{Remark}

Since the main concern of this paper is safety, we present next how to guarantee safety also during exploration, i.e., when the action applied to the system is not given by~\eqref{eq:robust_mpc}. Afterwards, we will further discuss open research questions and possible ways of addressing them.

\subsection{Safe Exploration}
The proposed formulation guarantees safety during exploitation, i.e., whenever the policy is given by~\eqref{eq:value_policy}. However, specific care needs to be taken during exploration, i.e., when the optimal policy is perturbed, since also in this phase constraint satisfaction must not be jeopardized. Exploration is typically performed by picking a random action among all feasible actions with a given probability distribution. The main difficulty in enforcing safety is related to the complexity of the set of safe actions 
\begin{align*}
\vect{A}(\vect{s}_0):=\{\, \vect{a}_0 \,|\, \exists \, \vect{a}_1,\ldots \, \text{s.t.} \, C\vect{s}_k + D\vect{a}_k + \vect{\bar c} \leq 0, \, \forall \, \vect{w}\in \mathcal{\bar W} \,\}.
\end{align*}
Since this set is implicitly approximated within robust MPC, 
this issue can be tackled by using a modified version of the robust MPC Problem~\eqref{eq:robust_mpc}:
\begin{subequations}
	\label{eq:exploration}
	\begin{align}
	\min_{\vect{x},\vect{u}} \ \ & \eqref{eq:robust_mpc_cost} + f\left (\vect{u}_0,\vect{q}\right ) \\
	\mathrm{s.t.} \ \ & \vect{x}_0 = \vect{s}, \eqref{eq:robust_mpc_dyn}, \eqref{eq:robust_mpc_pc}, \eqref{eq:robust_mpc_tc},
	\end{align}
\end{subequations}
with either $f\left (\vect{u}_0,\vect{q}\right ):= \rho\| \vect{u}_0 - \vect{q} \|$, or $f\left (\vect{u}_0,\vect{q}\right ):= \vect{q}^\top \vect{u}_0$, and a randomly chosen $\vect{q}$. Note that performing constrained exploration in policy gradient methods can produce a biased gradient estimation. Simple strategies to tackle this issue are proposed in~\cite{Gros2019e}.

\begin{theorem}
	Consider a convergent RL scheme solving Problem~\eqref{eq:safe_rl_sampled} where (a) robust MPC~\eqref{eq:robust_mpc} is used as function approximator; (b) available data is handled as detailed in Section~\ref{sec:data}; and (c) exploration is performed according to~\eqref{eq:exploration}. Assume that all new data yields $\vect{w}_k\in \mathcal{\bar W}$. Then, the RL scheme is safe in the sense of Definition~\ref{def:safety}, i.e., $\mathbb{P}\left [\exists \, k \, \mathrm{s.t.} \,\vect{\xi}(\vect{s}_k,\vect{a}_k) >0\right ]\leq 1-\eta(\mathcal{D})$. Additionally, the scheme is optimal in the sense of Definition~\ref{def:data_compression}. 
\end{theorem}
\begin{IEEEproof}
	We observe that $\eta\left(\mathcal D\right)$ quantifies the probability that it is possible that a scenario is unaccounted for in the construction of $\vect{\hat S}_+$. Moreover, we observe that if $\vect{\hat S}_+$ does not account for all possible scenarios, then $\vect{w}_k\notin \mathcal{\bar W}$ \textit{can} happen for some $k$. As a result, one can verify that 
\begin{align*}
\mathbb{P}\left [\exists \, k \, \mathrm{s.t.} \, \vect{w}_k\notin \mathcal{\bar W}\right ] \leq 1-\eta(\mathcal{D}).
\end{align*}	
	We note that, by Proposition~\ref{prop:feas_stab}, $\vect{w}_k\in \mathcal{\bar W}$ implies that robust MPC~\eqref{eq:robust_mpc} is recursively feasible. Moreover, exploration is performed using~\eqref{eq:exploration}, i.e.,~\eqref{eq:robust_mpc} with a modified cost, such that recursive feasibility is also preserved by Proposition~\ref{prop:feas_stab}. This entails that a constraint violation can only occur if $\vect{w}_k\notin \mathcal{\bar W}$. The converse, however, is not true in general, since there might exist $\vect{w}_k\notin \mathcal{\bar W}$ which does not entail a constraint violation. Consequently, the probability that the constraints be violated at any time is bounded from above by $1-\eta(\mathcal{D})$:
	\begin{align*}
		\mathbb{P}\left [\exists \, k \, \mathrm{s.t.} \,\vect{\xi}(\vect{s}_k,\vect{a}_k) >0\right ] \leq \mathbb{P}\left [\exists \, k \, \mathrm{s.t.} \, \vect{w}_k\notin \mathcal{\bar W}\right ] \leq 1-\eta(\mathcal{D}).
	\end{align*}
	Data compression optimality (Definition~\ref{def:data_compression}) is a direct consequence of Theorem~\ref{thm:efficiency}.
\end{IEEEproof}
	\begin{Remark}
		We ought to stress that assuming that RL converges to a minimum of $\mathbb{E}[\psi]$ is standard. However, assuming convergence to a local minimum of $J$ can be a strong assumption, which is typically not met by $Q$-learning and SARSA, while actor-critic methods are typically expected to converge to a local minimum of $J$ for the given parametrization. 
		Optimality in the sense of Definition~\ref{def:sdc_optimality} (set membership optimality) could then be claimed for convergent actor-critic methods, while it is reasonable to expect a certain degree of suboptimality with $Q$-learning.
	\end{Remark}

	\begin{algorithm}[t]
			\caption{RL MPC}\label{alg:rl_mpc}
			\begin{algorithmic}[1]
				\If{ Explore }
				\State Get $\vect{a}$ from MPC~\eqref{eq:exploration} \label{alg:mpc_e}
				\Else 
				\State Get $\vect{\pi}_{\vect{\theta}}(\vect{s})$ from MPC~\eqref{eq:robust_mpc} \label{alg:mpc}
				\EndIf
				\State Observe $\vect{s},\vect{a}\to \vect{s}_+$
				\State Compute $\vect{w}$, solve~\eqref{eq:chull_lp} to update $\mathcal{\bar W}$
				\State Get $Q_{\vect{\theta}}(\vect{s},\vect{a}), \nabla_{\vect{\theta}} Q_{\vect{\theta}}(\vect{s},\vect{a})$, $V_{\vect{\theta}}(\vect{s}_+)$
				\If{ Solve RL } \label{alg:solve_rl}
				\State Compute RL step: e.g., by solving~\eqref{eq:safe_rl_sampled} \label{alg:rl_step}
				\EndIf
				\If{ Update $\vect{\theta}$ feasible } \label{alg:update_theta}
				\State Recompute constraint tightening~\eqref{eq:constr_tightening},~\eqref{eq:terminal_constr_tightening} \label{alg:constr_tightening}
				\EndIf
			\end{algorithmic}
	\end{algorithm}

		We provide an overview of the computations performed by RL-MPC in Algorithm~\ref{alg:rl_mpc}. Note that we introduced lines~\ref{alg:solve_rl} and~\ref{alg:update_theta} since the RL problem and, consequently, the recomputation of the constraint tightening and terminal set might in principle be updated less often than the feedback sampling time. In that case, a batch RL approach could be used instead of a recursive one and the real-time requirements would only concern MPC. 
		Additionally, we stress that in our Matlab implementation which was only partially made efficient, the constraint tightening procedure on line~\ref{alg:constr_tightening} required approximately twice the computation time of MPC on lines~\ref{alg:mpc_e} and~\ref{alg:mpc}. The computation of the RL step on line~\ref{alg:rl_step} was approximately $5$ times faster.

\subsection{Discussion on the Proposed Approach}
\label{sec:discussion}

We discuss next a few open research questions and comment on possible extensions to the proposed framework.

\subsubsection{Safety}
Our safety definition is based on the assumption that all future samples satisfy $\vect{w}_k\in \mathcal{\bar W}$.
In practice this assumption, though standard in robust MPC and SMSI, can be rather strong. However, this is a fundamental problem of safety: one can never guarantee a priori that a bounded set contains all possible future realizations of an unknown stochastic process. In Section~\ref{sec:slack} we have proposed a simple and practical approach to retain MPC feasibility even in the case $\vect{w}_k \notin \mathcal{\bar W}$. However, with this approach safety is potentially lost every time a sample falls out of the convex hull $\mathcal{\bar W}$. Safety, however, is typically quickly recovered, as the RL parameter is instantaneously adjusted to account for the new sample. As also highlighted in Section~\ref{sec:safe_rl} and Remark~\ref{rem:violation} this temporary loss of safety is a fundamental issue for any sample-based approach, unless additional assumptions are introduced. 
While one could envision the derivation of some measure of reliability of the identified set to be used to provide stronger guarantees, such investigation is beyond the scope of this paper and will be the subject of future research. 

\subsubsection{Approximation Quality}
It has been proven in~\cite[Theorem 1]{Gros2018} that, provided that the MPC parametrization is rich enough, the correct value and action-value functions can be recovered exactly. In the context of this paper, however, the parametrization of the noise set is approximate by construction. Consider partitioning the parameter vector as $\vect{\theta}=(\vect{\theta}_\mathrm{c},\vect{\theta}_{\vect{W}})$, where $\vect{\theta}_\mathrm{c}$ is the parameter vector directly defining the cost and constraint functions, similarly to the parametrization of~\cite{Gros2018}, while $\vect{\theta}_{\vect{W}}$ is the parameter vector parametrizing function $\vect{g}_{\vect{\theta}}:=\vect{g}_{\vect{\theta}_{\vect{W}}}$. Then, provided that the parametrization of the cost and constraint functions through $\vect{\theta}_\mathrm{c}$ is rich enough, the value, action-value functions and policy can be recovered exactly on the feasible domain of the robust MPC~\eqref{eq:robust_mpc}. Therefore, as opposed to the result of~\cite{Gros2018}, the equivalence only holds on the subset of the feasible domain of the RL problem which can be described by the chosen parametrization of $\vect{g}_{\vect{\theta}}$.

Since the hypothesis of a perfect parametrization is unrealistic in most relevant applications, $Q$ learning and SARSA will in general not deliver the best performance that can be attained with the selected parametrization, since these algorithms aim at fitting the action-value function rather than directly optimizing performance. It is therefore appealing to resort to policy gradient approaches, which seek the direct minimization of $J$ by manipluating $\vect\theta$.
While in principle the proposed approach is well suited for policy gradient methods (both stochastic and deterministic), the main difficulty that needs extra care to be handled is related to the exploration strategy to be deployed, which in general introduces a bias in the gradients, such that convergence to a local optimum is hindered. This topic is investigated in~\cite{Gros2019e}.

\subsubsection{Model Adaptation}

In principle, one could choose to let RL update any of the parameters~\eqref{eq:mpc_parameters} of the MPC scheme~\eqref{eq:robust_mpc}, including the model parameters $A$, $B$, $\vect{b}$. The model used by MPC plays an important role in (a) the definition of the value and action-value function and (b) the conservatism of the uncertainty set approximation $\vect{W}_{\vect{\theta}}$. 
	As observed in Remark~\ref{rem:model_adaptation}, letting RL adapt the model parameters results in a large increase of computations.
	The feedback matrix $K$, however, can be adapted by RL without major issues. Since it has an impact on the size of the uncertainty propagation and, through $\vect{c}_k$ and $\vect{g}$, on the cost, letting RL adapt $K$ is expected to further reduce the closed-loop cost by reducing the tightening of the specific constraint components $\vect{c}_{k,i}$, $\vect{g}_j$ corresponding to constraints, which are active in the execution of the the control task. Note that selecting a $K$ that is optimal for the given task is an open issue which can be addressed in a data-driven fashion using the approach we propose.
	
	In~\cite[Theorem 1]{Gros2018} and~\cite[Corollary 2]{Zanon2019} it is proven that RL can find $Q_{\vect{\theta}}=Q$, $V_{\vect{\theta}}=V$ and $\vect{\pi}_{\vect{\theta}}=\vect{\pi}$ even without adjusting the model, provided that the parametrization is descriptive enough. However, typically the parametrization is low-dimensional and cannot be expected to fulfill this requirement.
	Therefore, the question on whether adapting the model provides an advantage and on how to best adapt it is still open. 

	One possibility to improve the approximation quality could be to carry a fixed model parametrization $A_0$, $B_0$, $\vect{b}_0$ to be used for safety and one or several (possibly nonlinear) models $\vect{f}_{i,\vect{\theta}}(\vect{x}_i,\vect{u}_i)$, each associated with its cost to construct a more accurate prediction of the future cost distribution in a scenario-tree fashion. The investigation of alternative formulations will be the subject of future research. 

\subsubsection{Stability}
The results on robust MPC are stronger than the one provided in Proposition~\ref{prop:feas_stab}: asymptotic stability  is proven, under the assumptions of having $\gamma=1$ and a positive-definite stage cost, usually called \emph{tracking} cost, as opposed to \emph{economic} cost.
In case of a tracking cost, the presence of a discount factor in the MPC formulation complicates the stability analysis. While one could foresee formulating~\eqref{eq:robust_mpc} with $\gamma=1$, even though $\gamma<1$ in the RL problem, it is not immediately clear to which extent this inconsistency will impact on the ability to learn the correct policy. While providing approximation guarantees for $Q$-learning is arguably nontrivial, policy gradient methods will at least provide the best policy for the selected parametrization. In this context that would be be best policy with stability guarantees. 
By exploiting the results of~\cite{Zanon2014d} it should be possible to prove that stabilizing linear control laws can be recovered exactly for linear systems. 
However, a thorough investigation of this topic is the subject of ongoing research.

The distinction between tracking and economic cost is relevant in relation to the stability properties of the closed-loop system: a thorough discussion on economic MPC can be found in~\cite{Diehl2011,Amrit2011a,Mueller2015}. The linear-quadratic case, which is particularly relevant for our framework, has been analyzed in, e.g.,~\cite{Zanon2014d,Zanon2016b,Zanon2018a,Faulwasser2018,Gruene2018}. We only recall here that all the conclusions drawn in~\cite{Gros2018} apply directly to our setup, i.e., the economic case can be reduced to the tracking case by additionally learning an initial cost.

\section{Simulation Results}
\label{sec:simulations}

We test our approach with two examples in simulations. First we consider a linear system to easily interpret the results. Then we consider a realistic example from the chemical industry: a nonlinear evaporation process.

\subsection{Linear System}
We consider a linear system having $2$ states, such that we can easily visualize the behavior of RL-MPC. Consider a simple linear system with dynamics and stage cost
\begin{align*}
	\vect{s}_+ &= \matr{cc}{1 & 0.1 \\ 0 & 1} \vect{s} + \matr{c}{0.05 \\ 0.1} a + \vect{w}, \\
	\ell(\vect{s},\vect{a}) &= \matr{c}{\vect{s}-\vect{s}^\mathrm{r} \\ a -a^\mathrm{r} }^\top \mathrm{diag}\left ( \matr{c}{1 \\ 0.01 \\ 0.01} \right ) \matr{c}{\vect{s}-\vect{s}^\mathrm{r} \\ a -a^\mathrm{r} },
\end{align*}
where $\vect{s}=(p,v)$.
We formulate a problem with prediction horizon $N=20$ and introduce the state and control constraints $-\vect{1}\leq \vect{s}\leq \vect{1}$, $-10\leq a\leq 10$.
The real noise set is selected as a regular octagon. 
In order to illustrate the ability of RL to adapt the approximation of the noise set, we select to parametrize $\vect{W}_{\vect{\theta}}$ as a polytope with 4 facets. 
We compute the terminal cost matrix $P$ and feedback gain $K$ using the LQR formulation resulting from the nominal model and stage cost. The terminal feedback $K$ is used for constraint tightening, as per Section~\ref{sec:rmpc}.

We simulate the RL-MPC scheme over $200$ time steps in a scenario without exploration in which the reference is 
\begin{align*}
	p^\mathrm{r}(t) &= \left \{\begin{array}{ll}
	1 & 25\leq t \leq 120 \\
	-1 & \text{otherwise}
	\end{array}\right., \\
	v^\mathrm{r}(t)&=0, && a^\mathrm{r}(t)=0.
\end{align*}
Since the setpoint reference is moving, for simplicity we impose the terminal constraint centered around the origin. While this does not affect recursive feasibility, practical stability is harder to prove in this case. A thorough analysis of this aspect goes beyond the scope of this paper, and we simply recall that such a terminal constraint induces a leaving arc in the MPC optimal control problem. This situation has been analyzed in~\cite{Faulwasser2018,Zanon2018a,Grune2013a} in the context of economic MPC, concluding that under suitable assumptions verified by the system considered here, practical stability is obtained. 

We update $\vect{\theta}=(M,\vect{m})$ according to~\eqref{eq:rl_param_update}-\eqref{eq:safe_rl_sampled} with $\alpha=0.1$, $\psi$ given by~\eqref{eq:q_learning} ($Q$ learning). 
Problem~\eqref{eq:safe_rl_sampled} is solved to full convergence at each step such that: (a) globalization techniques guarantee that $\vect{\theta}^*$ is a better fit than $\vect{\theta}_k$ for the sample at hand; (b) positive-definiteness of the cost yields a well-posed MPC formulation; and (c) the choice of parameter $\alpha$ is directly related to the horizon of a moving-average approximation of the expected value. 

\begin{figure}[t]
	\begin{center}
		\includegraphics[width=0.9\linewidth,clip,trim= 0 0 0 30]{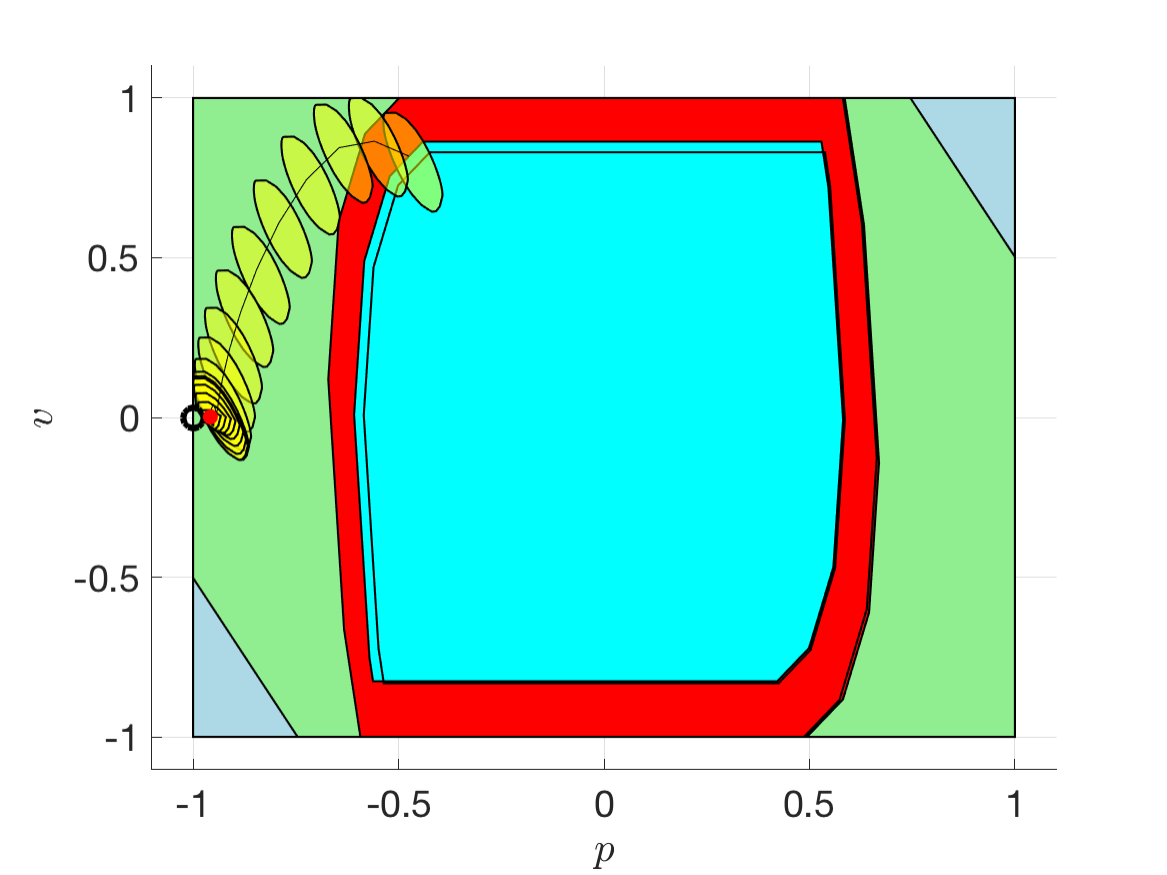}
		\includegraphics[width=0.9\linewidth]{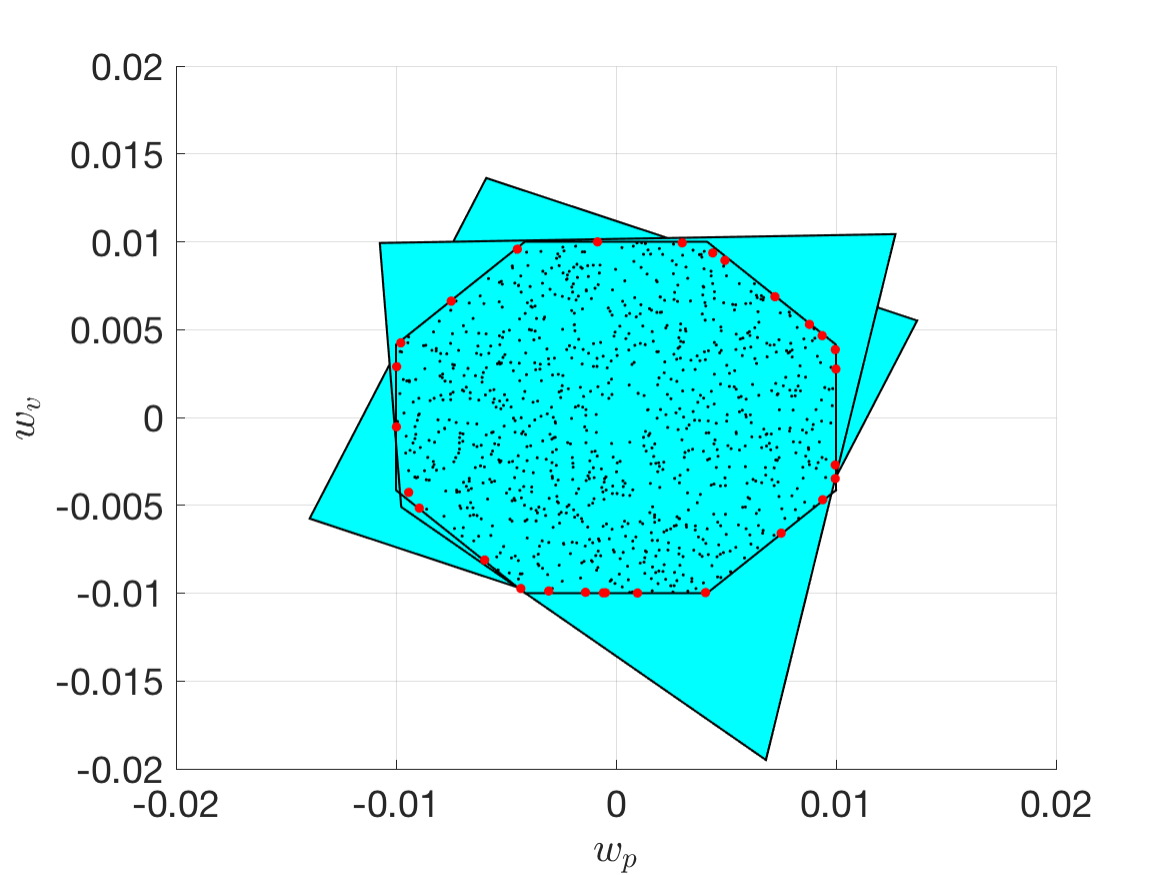}
	\end{center}
	\caption{Snapshots at $k=0$ and $k=24$. Top figure (state space): state constraint set (light blue), state-input constraints using feedback matrix $K$ (green), RPI set (red), terminal constraint set $\mathcal{X}_\mathrm{f}$ (cyan for $k=24$, transparent for $k=0$). Predicted trajectory at $k=24$: initial state (red dot), predicted trajectory (solid black line), reference $\vect{s}^\mathrm{r}$ (black circle), uncertainty tube (yellow). Bottom figure (noise space): true uncertainty set (transparent octagon), noise samples (black dots), vertices of their convex hull (red dots), uncertainty set approximations $\vect{W}_{\vect{\theta}}$ (cyan sets, with $k=0$ in the background). A better approximation  $\vect{W}_{\vect{\theta}}$ ($k=24$) enlarges $\mathcal{X}_\mathrm{f}$.}
	\label{fig:1_25}
\end{figure}

\begin{figure}[t]
	\begin{center}
		\includegraphics[width=0.9\linewidth,clip,trim= 0 0 0 30]{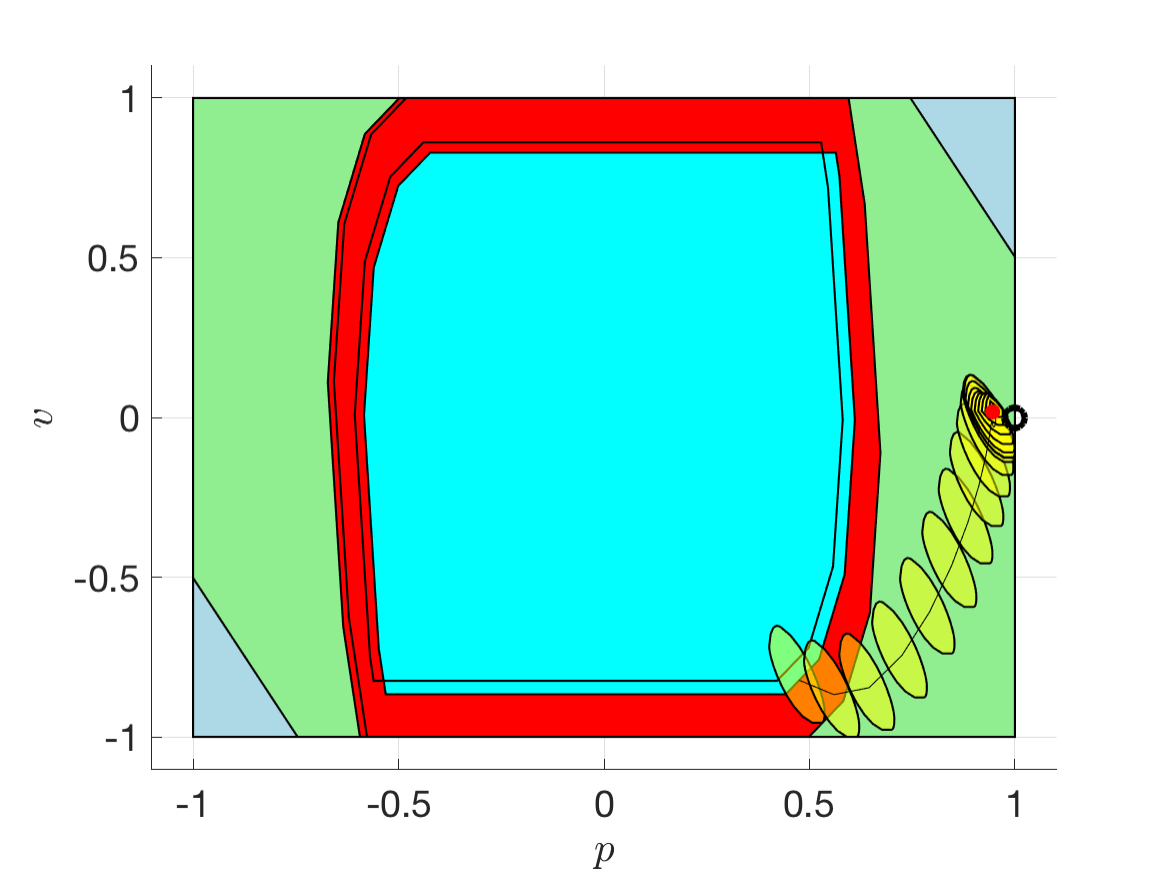}
		\includegraphics[width=0.9\linewidth]{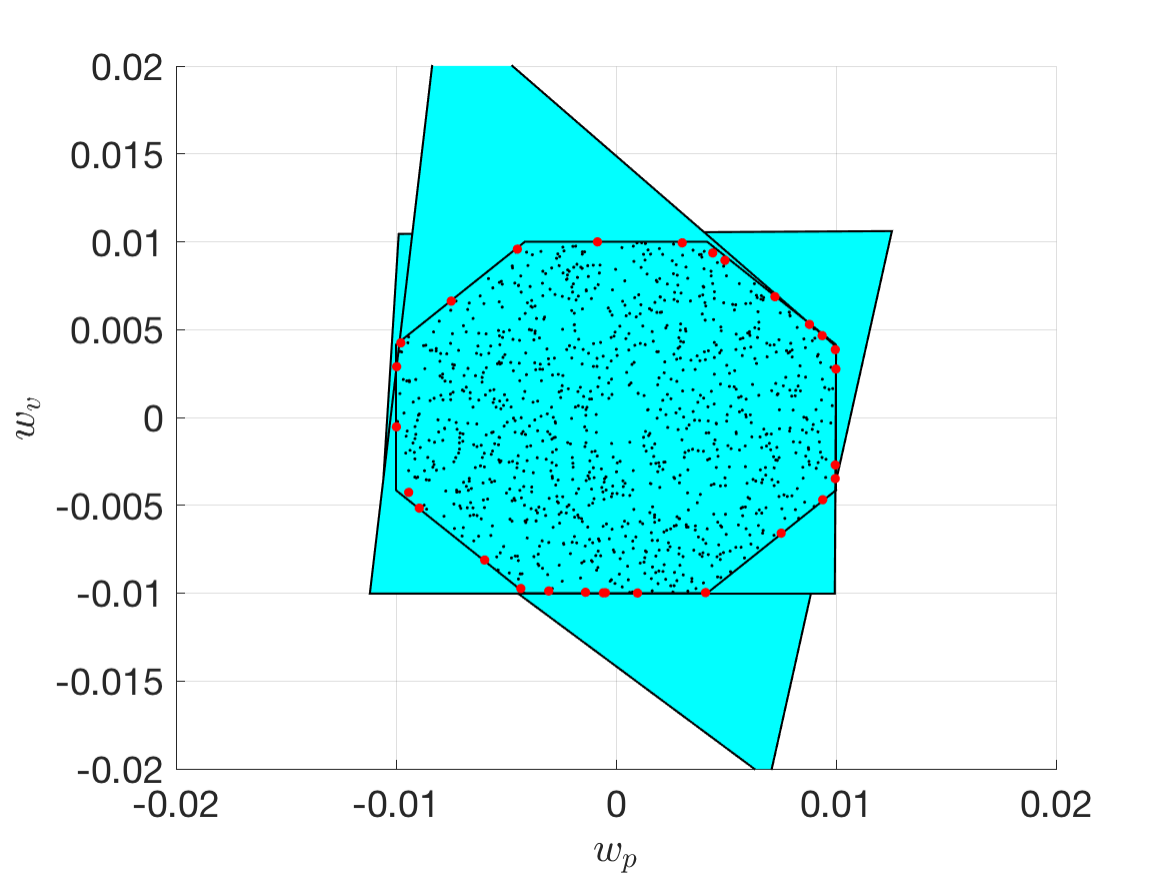}
	\end{center}
	\caption{Snapshots at $k=34$ and $k=109$: same convention as Figure~\ref{fig:1_25}, with predicted trajectory at $k=109$. Both the RPI and terminal sets moved closer to the setpoint by a better approximation $\vect{W}_{\vect{\theta}}$ for the specific control task (see the bottom plot). Moreover, next to $\vect{s}^\mathrm{r}$ the constraints are tightened more at $k=36$ than afterwards.}
	\label{fig:35_110}
\end{figure}

The simulation results are displayed in Figures~\ref{fig:1_25} and~\ref{fig:35_110} in the form of snapshots comparing a representation of the solver at two different time instants. In particular, the constraint, RPI and terminal sets are represented together with the trajectories and constraint tightening. Associated with these quantities, we also display the uncertainty set with the drawn samples, their convex hull and the approximation that RL learned. Initially, RL adapts the rather conservative approximation, therefore enlarging the terminal set by reducing the required constraint tightening. When the setpoint moves, there is initially no gain in modifying the set adaptation, until at $k=34$ the tightened constraint $p\leq1$ becomes active. At this point, RL starts adapting the set approximation to better capture the shape of the uncertainty set in its top-right part, which is opposite to the bottom-left corner which was better approximated before. Consequently, the terminal set is shifted towards the reference and the $p\leq1$ is tightened less, with an infinite-horizon difference of approximately $0.019$. Since the tightening steady-state is quickly reached, this difference is visible in Figure~\ref{fig:35_110}. 

Note that $\vect{m}$ does in principle not need to be adjusted, as any adjustment in $\vect{m}$ can be equivalently obtained by suitably rescaling $M$. We performed the same simulations with $\theta=M$ and obtained qualitatively equivalent results, the detailed presentation of which we therefore omit. Finally, note that the convex hull of $\mathcal{W}$ is a polytope with $28$ facets, as opposed to the used approximation $\vect{W}_{\vect{\theta}}$, which only has $4$ facets.

We ran an additional simulation in which we let $\vect{\theta}=(M,K)$, such that RL also adapts the feedback matrix $K$ used for constraint tightening. We remark that the problem of designing the terminal feedback and corresponding set $\mathcal{X}_\infty$ is nontrivial and many approaches have been developed. However, to the best of the authors' knowledge, none of these approaches explicitly accounts for the specific control task to be executed and rather aim at minimizing constraint tightening or maximizing the size of the RPI set. 
The simulation results are similar to the previous case. However, RL acts on $K$ to enlarge the RPI and terminal set while reducing the required constraint tightening, therefore obtaining an increase in closed-loop performance.

\subsection{Evaporation Process}

Consider the evaporation process modeled in~\cite{Wang1994,Sonntag2006} and used in~\cite{Amrit2013a,Zanon2016b} to demonstrate the potential of economic MPC in the nominal case. 
The model has states $\vect{x} = (X_2, \, P_2)$ (concentration and pressure); controls $\vect{u}=(P_{100},\,F_{200})$ (pressure and flow); and dynamics given by~\cite{Amrit2013a}
\begin{align}
M \dot X_2 &= F_1 X_1 - F_2 X_2, &
C \dot P_2 &= F_4-F_5,
\end{align}
where 
$T_2 = aP_2 + bX_2 + c$, $T_3 = dP_2 + e$, $\lambda F_4 = Q_{100} - F_1C_\mathrm{p}(T_2-T_1)$, $T_{100} = fP_{100} + g$, $Q_{100} = UA_1(T_{100}-T_2)$, $UA_1 = h(F_1+F_3)$, $Q_{100} = UA_1(T_{100}-T_2)$, $UA_1 = h(F_1+F_3)$, $Q_{200} = \frac{UA_2(T_3-T_{200})}{1+UA_2/(2C_\mathrm{p}F_{200})}$, $F_{100} =\frac{Q_{100}}{\lambda_\mathrm{s}}$, $\lambda F_5 = Q_{200}$, $F_2 = F_1-F_4$.
The model parameters are given in~\cite{Zanon2016b}. 

Concentration $X_1$, flow $F_2$, and temperatures $T_1,T_{200}$ are stochastic. In this example, we model them as a uniform distribution with $X_1=5\pm0.5$, $F_1=10\pm0.5$, $T_1=40\pm4$, $T_{200}=25\pm5$. Additionally, the controller must satisfy bounds $(25,40)\leq (X_2,P_2) \leq (100,80)$ on the states and $100 \leq (P_{100},F_{200}) \leq 400$ on the controls. 
The stage cost is given by 
\begin{align*}
\ell(\vect{x},\vect{u}) =\ &10.09(F_2+F_3) + 600 F_{100} + 0.6 F_{200},
\end{align*}
which entails that the nominal model is optimally operated at the steady state $\vect{x}_\mathrm{s}=(25,49.74)$, $\vect{u}_\mathrm{s}=(191.71,215.89)$, with stage cost $\ell(\vect{x}_\mathrm{s},\vect{u}_\mathrm{s}) =\ell_\mathrm{s}$.

The system is linearized at $\vect{x}_\mathrm{s},\vect{u}_\mathrm{s}$ to obtain a linear nominal model. The terminal set is centered at $\vect{x}_\mathrm{c}=(29,53.57)$, $\vect{u}_\mathrm{c}=(223.76,221.61)$, which is a steady state for the linearized dynamics. 
Since the safety constraints are already linear, we introduce discount factor $\gamma=0.99$ and formulate a robust linear MPC problem of the form~\eqref{eq:robust_mpc}. To that end, we use the linearization at $x_\mathrm{s}$, $u_\mathrm{s}$, the quadratic cost obtained by applying the tuning procedure proposed in~\cite{Zanon2016b}. With the given stage cost and linear model, we obtain the LQR feedback $K$ which we use to stabilize the model error in computing the constraint tightening~\eqref{eq:constr_tightening} and the terminal set~\eqref{eq:terminal_constr_tightening}. 
We then use RL to adjust parameter $\vect{\theta}=(\vect{h}, \vect{p},M,K)$, i.e., the cost gradient, the uncertainty set and the feedback matrix $K$. Since we formulate the robust MPC problem~\eqref{eq:robust_mpc} using a positive-definite cost, while the RL cost $\ell$ is economic, we also learn a quadratic initial cost, as briefly discussed in Section~\ref{sec:discussion} and fully justified in~\cite{Gros2018}.

\begin{figure}[t]
	\begin{center}
		\includegraphics[width=0.9\linewidth,clip,trim= 0 0 0 30]{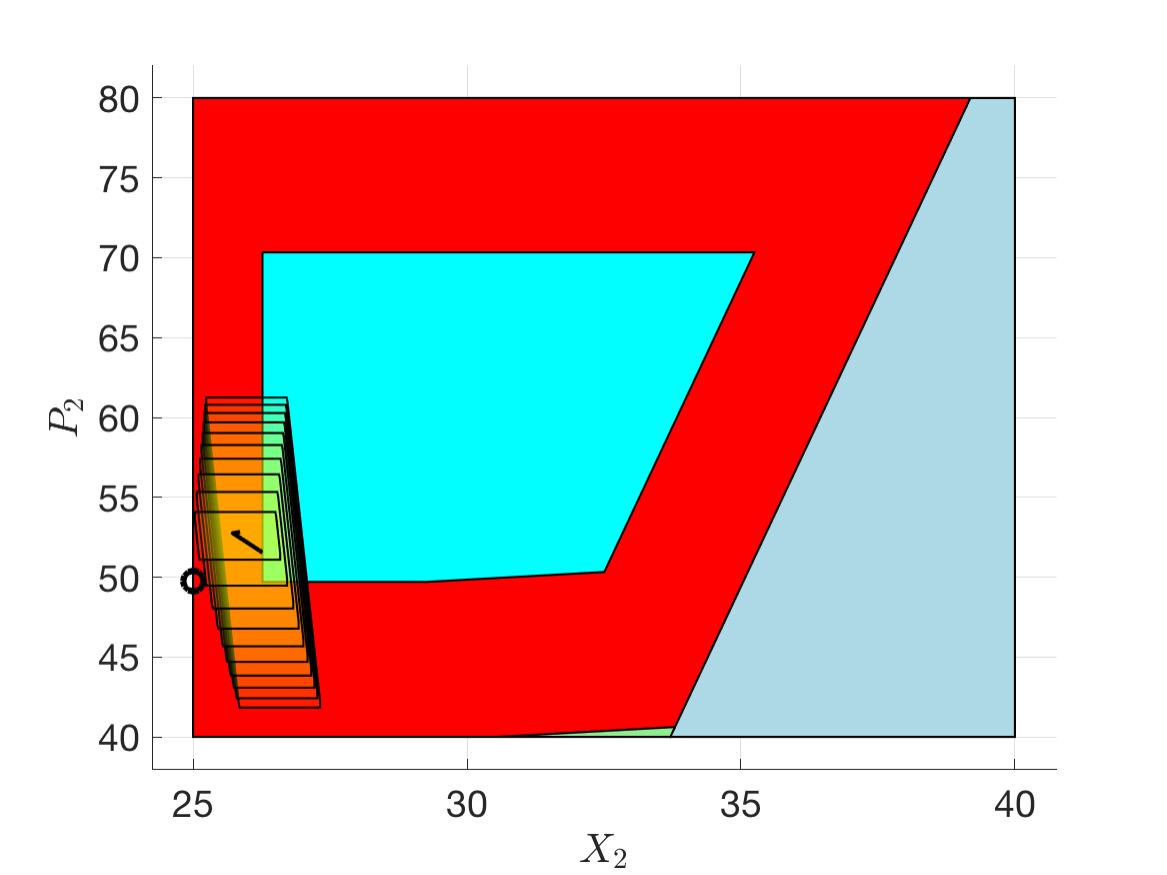}
		\includegraphics[width=0.9\linewidth,clip,trim= 0 0 0 30]{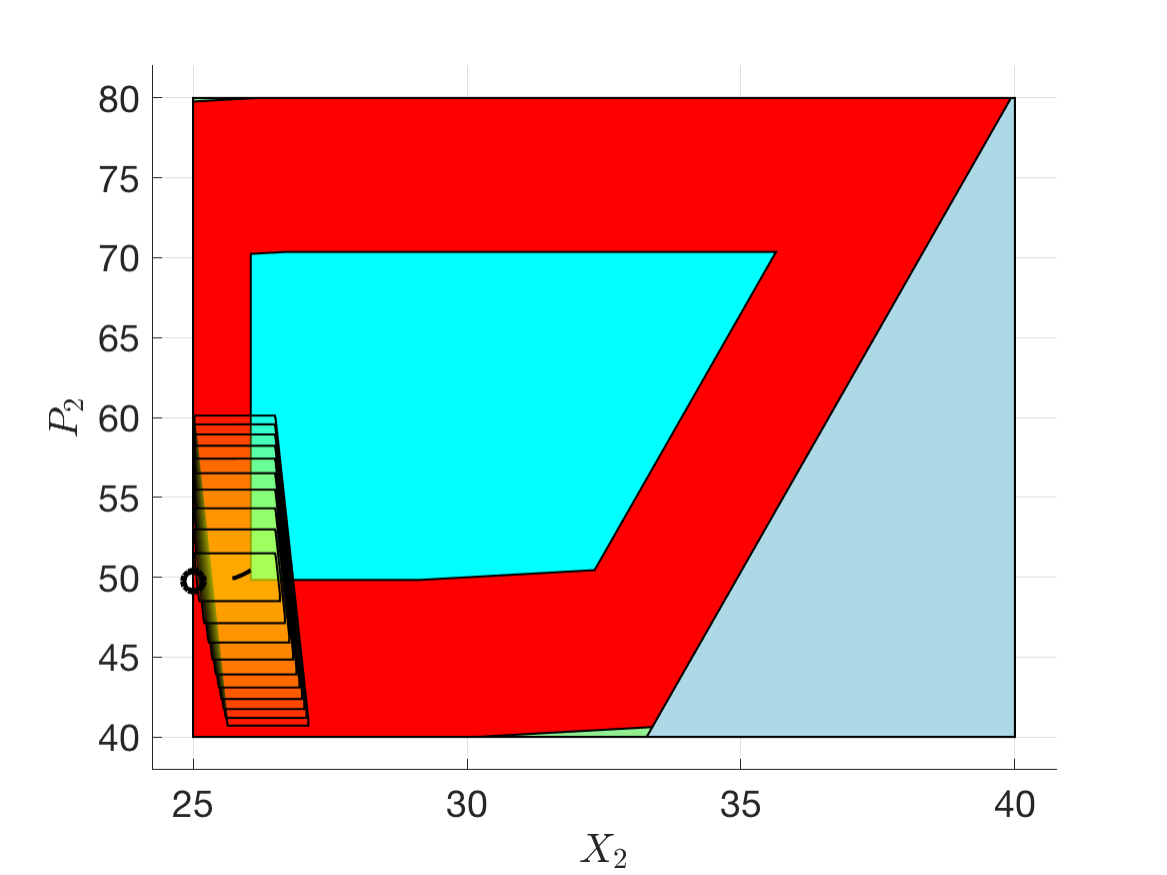}
		\includegraphics[width=0.9\linewidth,clip,trim= 0 0 0 30]{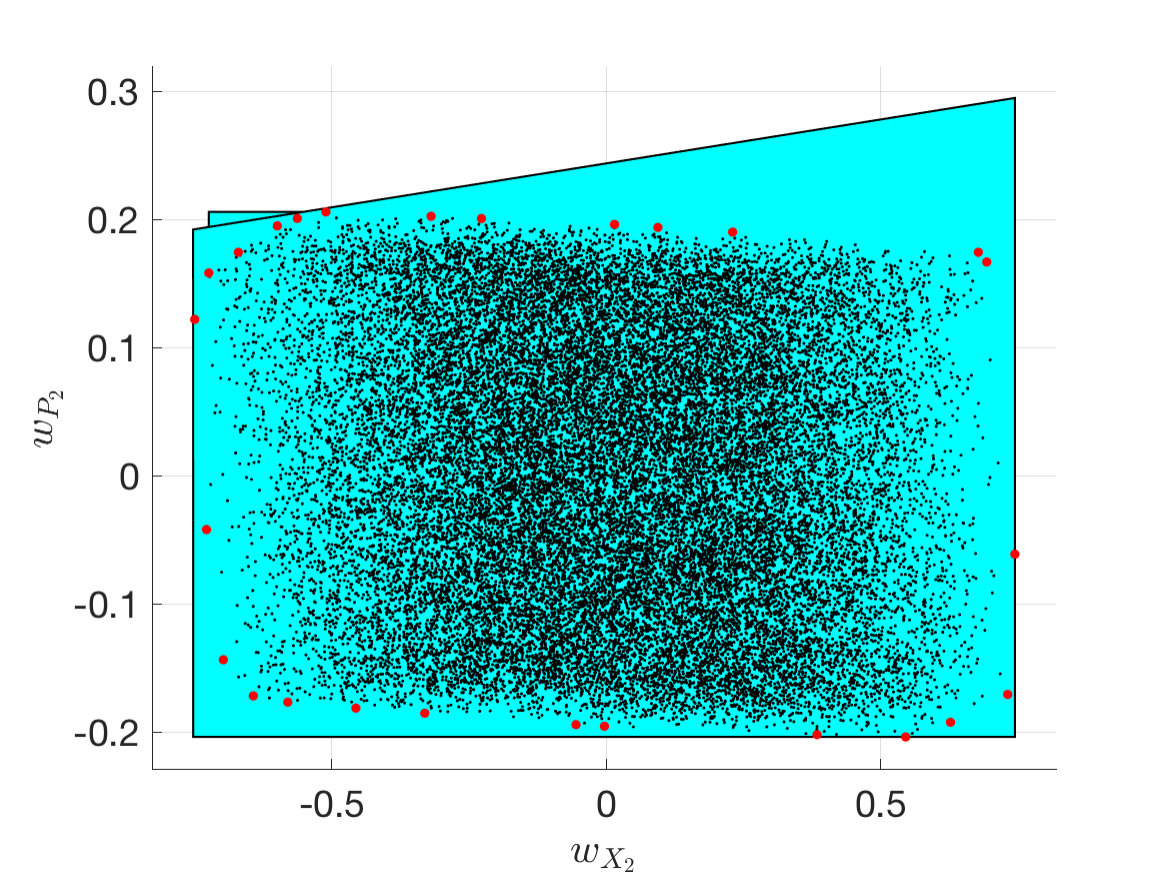}
	\end{center}
	\caption{Evaporation process. Top two figures: RPI and terminal set at the beginning and end of the learning process, same color convention as Figure~\ref{fig:1_25}. Bottom figure: the uncertainty set approximation $\vect{W}_{\vect{\theta}}$.}
	\label{fig:evap}
\end{figure}

We let the $Q$-learning algorithm run for $10^4$ samples with $\alpha = 10^{-2}$. We note that the parameters are no longer adjusted towards the end of the learning procedure, indicating convergence of the algorithm. 
The RPI and terminal set, as well as the uncertainty set approximation $\vect{W}_{\vect{\theta}}$ are displayed in Figure~\ref{fig:evap} at the beginning and at the end of the learning process. One can note that $\vect{W}_{\vect{\theta}}$ becomes larger in order to better approximate the top left part of the uncertainty set. In combination with the adjustment of the feedback matrix $K$, this allows to shift the terminal set towards the reference, as shown in the top plot.

\section{Conclusions and Future Work}
\label{sec:conclusions}

In this paper we have presented an RL algorithm which is guaranteed to be safe in the sense of strictly satisfying a set of prescribed constraints given the available data.
We have discussed both an innovative function approximation based on robust MPC and an efficient management of data which makes it possible to deal with very large amounts of data in real-time. 
While linear MPC can be successfully applied also to nonlinear systems, as demonstrated with the second example, in case of strong nonlinearities linear MPC might fail at providing satisfactory performance and even safety.

The proposed framework paves the road for several extensions: (a) one can easily foresee the use of robust MPC based on scenario-trees (which is also suitable for nonlinear models); (b) the use of stochastic or deterministic policy gradient is expected to further improve performance, as discussed in Section~\ref{sec:safe_rl}; (c) a formulation using the computational geometry approach for robustness and scenario trees for a refined cost approximation could be envisioned. 
These research directions are the subject of ongoing research.

\bibliographystyle{IEEEtran}
\bibliography{syscop}

\begin{IEEEbiography}[{\includegraphics[width=1in,height=1.25in,keepaspectratio]{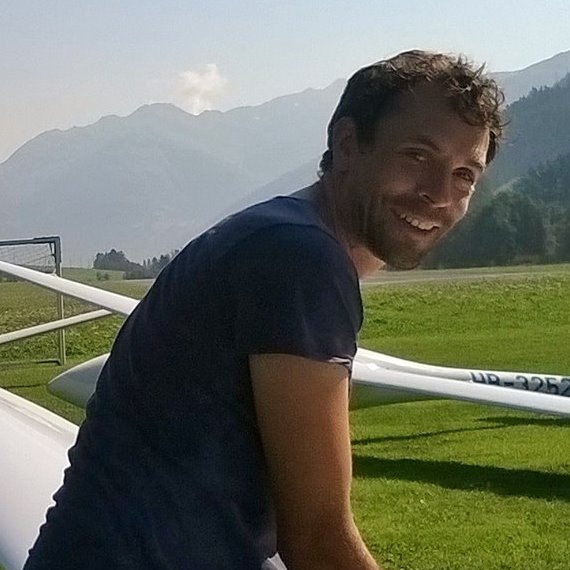}}]{S\'ebastien Gros}
	received his Ph.D degree from EPFL, Switzerland, in 2007. After a journey by bicycle from Switzerland to the Everest base camp in full autonomy, he joined a R\&D group hosted at Strathclyde University focusing on wind turbine control. In 2011, he joined the university of KU Leuven, where his main research focus was on optimal control and fast MPC for complex mechanical systems. He joined the Department of Signals and Systems at Chalmers University of Technology, G\"{o}teborg in 2013, where he became associate Prof. in 2017. He is now full Prof. at NTNU, Norway and guest Prof. at Chalmers. His main research interests include numerical methods, real-time optimal control, reinforcement learning, and the optimal control of energy-related applications.
\end{IEEEbiography}
\begin{IEEEbiography}[{\includegraphics[width=1in,height=1.25in,clip,keepaspectratio]{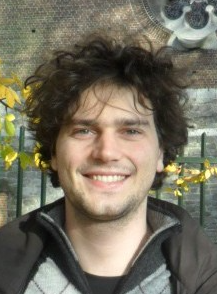}}]{Mario Zanon}
	received the Master's degree in Mechatronics from the University of Trento, and the Dipl\^{o}me d'Ing\'{e}nieur from the Ecole Centrale Paris, in 2010. After research stays at the KU Leuven, University of Bayreuth, Chalmers University, and the University of Freiburg he received the Ph.D. degree in Electrical Engineering from the KU Leuven in November 2015. He held a Post-Doc researcher position at Chalmers University until the end of 2017 and is now Assistant Professor at the IMT School for Advanced Studies Lucca. His research interests include numerical methods for optimization, economic MPC, optimal control and estimation of nonlinear dynamic systems, in particular for aerospace and automotive applications.
\end{IEEEbiography}

\end{document}